\documentclass[aps,pra,twocolumn,reprint,superscriptaddress]{revtex4-1}

\usepackage{amsmath,amssymb,graphicx,bbold}
\usepackage{graphicx}  % needed for figures
\usepackage{dcolumn}   % needed for some tables
\usepackage{bm}        % for math

\usepackage[caption=false,position=top,margin=12pt,justification=raggedright,singlelinecheck=false]{subfig}
\usepackage{cleveref}
\usepackage{amsmath}    % need for sub equations
\usepackage{graphics}
\usepackage{verbatim}   % useful for program listings
\usepackage{color}      % use if color is used in text
\usepackage{url}
\usepackage{breakurl}
\usepackage[breaklinks]{hyperref}   % use for hypertext links, including those to external documents and URLs
\hypersetup{
  colorlinks,
  citecolor=blue,
  linkcolor=blue,
  urlcolor=blue}
\usepackage{lineno} %number lines
\usepackage{textcomp}% use apostrophe
\begin{document}

\title{The statistics of chaos in a bursting laser}

\author{Wendson A. S. Barbosa}
\email{wendson.barbosa@ufpe.br}
\affiliation{Departamento de F\'{\i}sica,~Universidade Federal de Pernambuco, 50670-901, Recife, Brazil}

\author{Edison J. Rosero}
\affiliation{Departamento de F\'{\i}sica,~Universidade Federal de Pernambuco, 50670-901, Recife, Brazil}
\affiliation{Universidad del Pac\'ifico, Departamento de Ciencias Naturales y Exactas, Buenaventura, 764503, Colombia}

\author{Jorge R. Tredicce}
\affiliation{Departamento de F\'{\i}sica,~Universidade Federal de Pernambuco, 50670-901, Recife, Brazil}
\affiliation{Universit\'{e} de la Nouvelle Caledonie (UNC), Pole Pluridisciplinaire de la Matiere et de l'Environnement (PPME), Noumea, EA3325, Nouvelle Caledonie}

\author{Jos\'e R. Rios Leite}
\email{rios@df.ufpe.br}
\affiliation{Departamento de F\'{\i}sica,~Universidade Federal de Pernambuco, 50670-901, Recife, Brazil}

%\date{\today}

\begin{abstract}
We demonstrate experimentally how  semiconductor lasers subjected to double optical feedback change the statistics of their chaotic spiking dynamics from Gaussian to long-tail Power Law distributions associated to the emergency of bursting. These chaotic regimes, which are features of excitable complex systems, are quantified by the tail exponent $\alpha$ and appear by changing the ratio between the feedback times. Transitions to bursting occur in the neighbourhood of low order Farey fractions. The physics behind these transitions is related to the variation of threshold pump current in the compound system as obtained from a deterministic set of rate equations. Numerical integration also verifies the observed chaos transitions indicating the possibility of controlling the bursting chaotic statistics. 
\end{abstract}


%05.45.Ac	Low-dimensional chaos
%05.45.Pq	Numerical simulations of chaotic systems
%05.45.-a	         Nonlinear dynamics and chaos
%42.55.Px	Semiconductor lasers; laser diodes
%42.60.Mi	Dynamical laser instabilities; noisy laser behavior 
%42.65.Sf  	Dynamics of nonlinear optical systems; optical instabilities, optical chaos and complexity, and optical spatio-temporal dynamics

\maketitle

\section{Introduction}

Within optics, the non-linear dynamics of lasers has been studied for a long time,
both for fundamental \cite{Ott} and applied physics \cite{sciamanna-nature2015} interests. 
Measuring a laser output power is a well developed procedure to determine the amplitude squared of 
the dynamical variable which is the emitted electric field. Furthermore,  
the ease of changing parameters in the laboratory permits the observation of large dynamical diversity. 
Another advantage of dynamical studies in lasers arrangements is the existence of model equations to 
compare and predict experimental results. So, beyond their interest for themselves they offer examples 
for analogies in the understanding of others complex systems. Among relevant examples are the dynamical 
systems with time delayed feedback in biology, as described by Mackey-Glass equations \cite{mackey-glass1977-science}.

%Optical FB - LFF and Excitability

Excitability, which is an important concept in the study of complex transitory states like in neurons \cite{Rinzel1998,izki,Laing2003,NeuronBurstReview}, was also 
addressed in laser dynamics \cite{giudici,mos,romariz,hurtado,coomans,selmi,eguia2,jhonPRL2008}.
Delayed optical feedback has proved to induce excitable \cite{giudici,eguia1} and chaotic behavior \cite{kim,massoler1} on semiconductor lasers.
It is well known that these lasers with optical feedback 
generate chaos which, after an averaging process due to detector bandwidth, manifests with irregular 
sharp drops in its intensity, 
usually called low frequency fluctuation (LFF) \cite{Risch}. 
The statistical properties of these LFF pulses have been extensively investigated \cite{hohl,sukow,mulet,massoler2014}.

%Second Cavity
Two simultaneous different feedbacks are simple schemes to get bursting of power drops with statistics on-demand as we will show.  Indeed, the addition of a second time delay feedback on a laser was  already proposed and studied, using two external 
optical cavities \cite{rogisternum,rogisterexp,pisarchik,jia}. High dimensional
chaotic dynamic was attained experimentally using such configuration \cite{fisher}. With this scheme 
and varying the lengths of the cavities and the intensity of the feedbacks, chaos suppression was also 
achieved \cite{liu}. Regions with stable emission and periodical behavior were observed \cite{liu}, 
demonstrating the system high sensitivity to double feedback. More than two feedbacks has also been addressed in the literature \cite{tobens} aiming at laser stabilization.

%Our Exp 

Our experiments reported here were done in a semiconductor laser, 
chaotic by the effect of double optical feedback. 
Measured laser intensity time series are the essential data in our work. 
The statistical characterization of the chaotic dynamics is done extracting the inter-spike times 
represented in histograms and obtaining the so-called tail index ${\bf \alpha}$, 
for alpha-stable distributions.
It is shown how parameter tuning in the system makes 
the inter spike time go from chaotic with Gaussian 
statistics to chaos with heavy tail L\'evy distributions among pulse bursting.
To corroborate the experiments we also solved numerically the rate equations 
for a mono-mode semiconductor laser including more than one feedback time. 
We not only verify semi-quantitative experimental 
aspects of our results but we get 
evidence that deterministic non-linear equations with multiple feedback times 
can account for the production of bursting spike dynamics followed by switching 
between chaotic, periodic, and stable states. 

\section{Experimental Setup}

\begin{figure}[!h]  
\includegraphics[width=\linewidth]{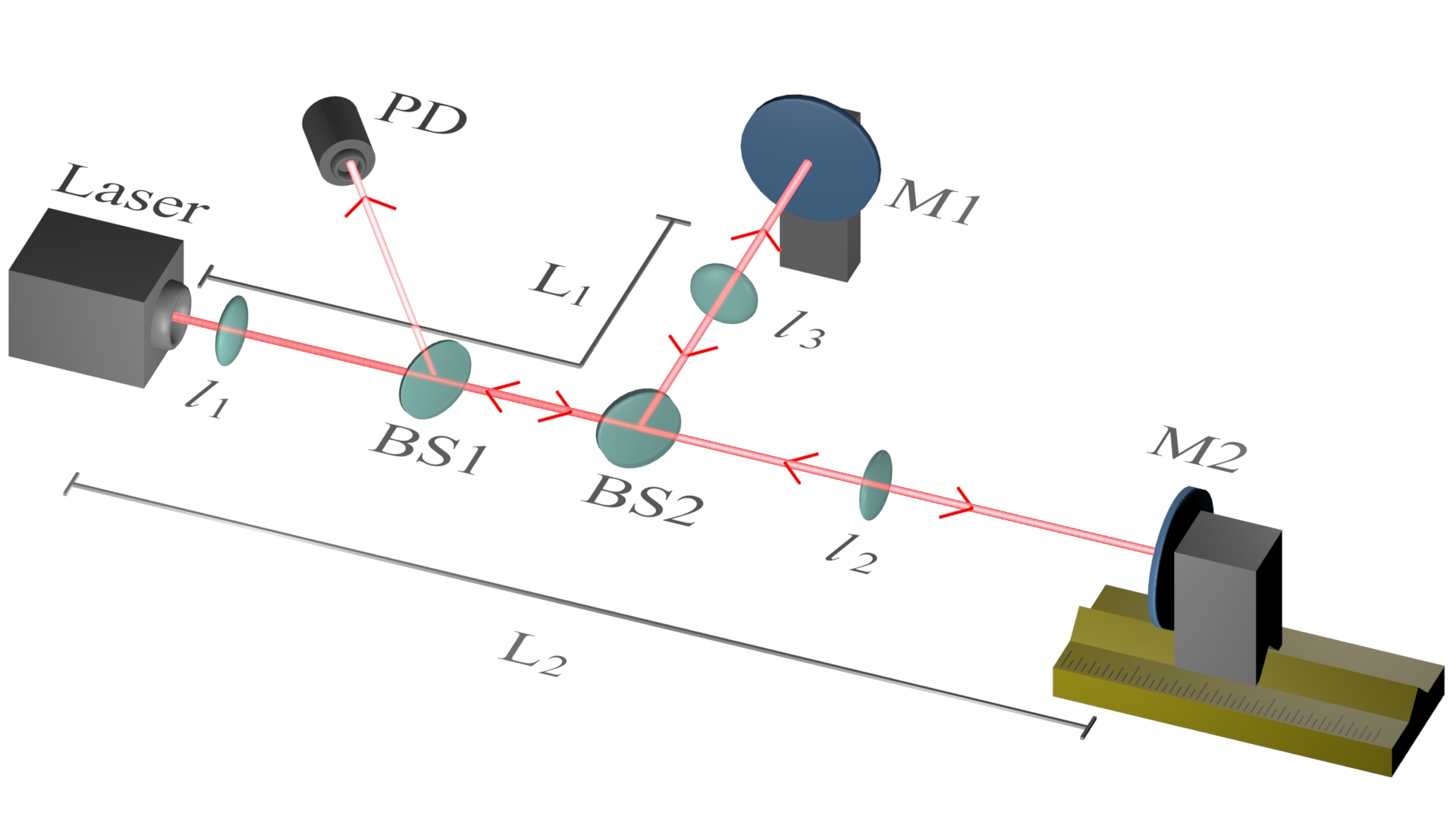} %0.55
\caption{Experimental setup. The laser output light is collimated by an aspheric lens $l_1$ and the data signal is measured by a 2.0 GHz bandwidth photodiode PD using a beam splitter BS1. The feedbacks are provided by two external cavities created by the beam splitter BS2 and the mirrors M1 (fixed position $L_1$) and M2 (variable position $L_2$). The lenses $l_2$ and $l_3$ are used to get optimum mode matching alignment condition in the two external arms.}
\label{Fig1}
\end{figure}

Our experimental data was obtained in the form of long time series for the fluctuating laser intensity obtained using the setup shown in Fig.~\ref{Fig1}. The experiments were done with commercial single mode diode lasers stabilized to 0.01 K, emitting in the infra-red wavelength and with solitary threshold current $J_{th}$ around 20 mA giving output power of few mW. 
Two specific units were SDL 5401 GaAlAs at 800nm - $J_{th}= 26$ mA and Thorlabs L850P010 at 850nm - $J_{th}= 10$ mA. % 26.66 mA
As indicated in Fig.~\ref{Fig1} the beam splitter BS1 allows the laser output 
intensity to be measured by a 2.0 GHz bandwidth photodiode PD and the data series 
collected with a Tektronix DPO 7104, 1 GHz oscilloscope. 
The two external cavities are created by the $50/50$ beam splitter BS2 and two mirrors $M_1$ and $M_2$. 
The fixed mirror $M_1$ create the first external cavity with a constant laser light pathway $L_1=8.02$ m corresponding to a fixed feedback 
time delay, $\tau_{1} = $53.5 ns, due to light round trip time. 
The second external cavity, created by the variable position mirror $M_2$, has a variable light 
pathway $L_2$, giving the second time delay, $\tau_2$, which is scanned during the experiments.
%
%
%%%%% FIGURE 1 - EXPERIMENTAL SETUP %%%%%
%

Experimental cavity alignment conditions were maintained using an aspheric collimating lens $l_1$
for beam divergence reduction.  Operation current was kept constant close to the solitary laser 
threshold, but up to 3\% current threshold reduction due to feedback could be achieved in both arms.  
Optimum alignment condition in the two external arms was imposed by inserting AR coated lenses $l_2$ 
and $l_3$ in front of each feedback mirror, in order to get mode matching. 
Consequently, maximum output intensity for lengths below 0.90 m and with one single feedback, remains 
in stationary conditions (without spikes). In this configuration, 
semiconductor laser with optical feedback exhibits physical properties of excitable systems when 
noise or any perturbation is applied \cite{giudici}. 
The best alignment was maintained during the variation of $L_{2}$ for lengths over 0.90 m. 

\section{Theoretical Model}

The dynamics of a semiconductor laser under feedback operation is well described by the so called 
Lang-Kobayashi equations \cite{LK}. This model has been previously used in order to include additional 
delayed term \cite{fisher,liu,rogisternum,pisarchik}. Here, we use the deterministic 
form of these equations for a system composed by two external cavities which provides a double feedback:
%
%%%%% LK-EQUATIONS - 2 FB %%%%%
%
\begin{eqnarray}
\frac{dE(t)}{dt}&=&\left[\frac{G_{N}-\gamma_p}{2}\right]E(t) + \sum_{i=1}^{2}\kappa_{i}E(t-\tau_{i})\cos{\Theta_i} \label{eq:LK2FB-E}\\ 
\frac{d\Phi(t)}{dt}&=&\frac{\overline{\alpha} \left[G_{N}-\gamma_p\right]}{2} -\sum_{i=1}^{2}\kappa_{i}\frac{E(t-\tau_{i})}{E(t)}\sin{\Theta_i} \label{eq:LK2FB-Phi}\\ 
\frac{dN(t)}{dt}&=& J - \frac{N(t)}{\tau_{s}} - G_{N}|E(t)|^{2} \label{eq:LK2FB-N}.
\end{eqnarray}\\
To be noticed is the absence of stochastic terms in the equations. 
Thus all random type dynamics results from the deterministic nonlinear nature of the model 
which reproduce the main dynamics of these kind of systems even to obtain chaotic behavior. 
Implementation of a fourth order Runge-Kutta routine with an integration step of $\delta t = 10^{-13}$\,s 
were done in C++. This value for $\delta t$ was shown to be small enough for the equations integrations 
during tests varying its value.
In the equations $i=1,2$ is the number of external cavities and 
$G_{N} = G_{0}\left[N(t)-N_0\right]/\left[1+\epsilon |E(t)|^{2}\right]$ is the medium gain. 
The optical frequency when the laser is solitary (with no feedback) is 
represented by $\omega_0$ and $\Theta_i =\omega_0 \tau_i + \phi(t) - \phi(t-\tau_i) $ 
is the field phase excursion after one external cavity round trip. 
The dynamical variables are the slow (with respect to the optical period) envelope 
for the electric field, having $E(t)$ as adimensional amplitude, 
$\Phi$(t) as the associated phase and $N(t)$ the carrier density. 
The parameter in the equations are: $\tau_{p}=1/\gamma_p$ and $\tau_{s}=1/\gamma_s$  
corresponding to photon and carrier lifetime, 
respectively; $G_{0}$ the small signal gain coefficient,  $N_{0}$ representing the 
carrier density at the transparency of the medium, $J$ is the injected pump current ($J_{th}$ 
represents threshold current), $\epsilon$ the saturation coefficient, $\kappa_{i}$ 
the feedback coefficients, $\tau_{i}$ the feedback delay times and $\overline{\alpha}$ the 
line width enhancement factor (not to be confused with the statistical index $\bf{\alpha}$).
The parameters values used were $\tau_s=2.0$ ns, $\tau_p=3.55$ ps, $\epsilon=5\times10^{-7}$, $G_0=3.2\times10^{3}$ s$^{-1}$, $N_0=1.5\times10^{8}$, $\kappa_1=\kappa_2=11\times10^{9}$ s$^{-1}$ and $\overline \alpha=3.0$.

\section{Statistics of Long-Tail Alpha-Stable L\'evy Distribution}

Observation of very different statistical distributions in our system naturally lead us to examine the well known alpha-stable type of distribution which present a good formalism to characterize shapes of distributions with respect  to their location, scale, distortion and tail \cite{LevyDistributions, McCulloch}. The experimental and numerical-theoretical data in our work were assumed to have stable distribution.
 
A general stable probability distribution, also known as L\'evy alpha-stable, stable Paretian or $\alpha$-stable distribution, is defined by the invariance property of a sum between two independent random variables with the index of stable distribution equal to $\alpha$ having as the result an $\alpha$-distribution with the same index. Often, the probability density function (PDF) $f(x)$ for a stable distribution does not have a general formula and some times it is not even analytical. The description of those distributions is better done by the characteristic function $\phi(t)$. The PDF, when it exists, is given by the Fourier transform of $\phi(t)$ which for stable distributions can be written as:
\begin{eqnarray}
\phi (t)= exp (it\delta-|\gamma t|^\alpha(1-i\beta sgn(t)\Phi)) 
\label{phi}
\end{eqnarray}
where i is the imaginary number, $sgn(t)$ is the signal of $t$ and $\Phi=-\frac{2}{\pi}\log{|t|}$ if $\alpha=1$, whereas $\Phi= \tan{ \left( \frac{\pi\alpha}{2}\right)}$ if $\alpha \neq 1$.
The stable parameter $\alpha$ is, for our data analysis, the most important among the four that characterize these distributions. It is also known as the tail index, index of stability or the characteristic exponent. Stable distributions have $0<\alpha \leq 2$ and it reveals the concentration of the distribution and its tail character. The upper limit, $\alpha=2$, correspond to the Gaussian case while long tail distributions have $\alpha<2$. Cauchy or Lorentz distributions have $\alpha=1$. Variance is undefined for $\alpha<2$ and mean does not exist for $\alpha \leq 1$. The others three parameters that caracterize an $\alpha$-stable distribution are related to the location $\delta$ $\in [-\infty,+\infty]$, to the scale $\gamma$ $\in [0,\infty]$ and to the skewness $\beta$ $\in [-1,+1]$. For the Gaussian case, i.e., $\alpha=2$ and $\beta=0$ due to the symmetry of this distribution, $f(x)$ can be obtained as it follows:
\begin{eqnarray}
f(x)=&\frac{1}{2\pi}\int_{-\infty}^{\infty} e^{-itx} \phi(t)dt = \frac{1}{\sqrt{2}\gamma \sqrt{2\pi}}e^{-\frac{(x-\delta)^2}{2(\sqrt{2}\gamma)^2}}.
\end{eqnarray}
A comparison with the well known PDF formula for Gaussian distribution allows us to relate its mean $\mu$ and variance $\sigma$ to the location and scale parameters $\delta=\mu$ and $\gamma = \sigma/\sqrt{2}$, respectively.
Once we had histograms of probability distributions for time interval between intensity spikes, we estimated $\alpha$ directly from this data using the software developed by Nolan \cite{NOLAN}. It calculates the four parameters for stable-like distribution with an algorithm is based on the quantile fit process of McCulloch \cite{McCulloch}. In our data, the occurrence of spike-bursting causes the deviation of $\alpha$ from the Gaussian condition to the long-tail distributions with $\alpha$ much close to 1 than 2.

\section{Experimental and Theoretical Results}

\subsection{Chaos with Single Feedback}

\begin{figure}[!h]
\centering
	\subfloat{
	\includegraphics[width=0.7\linewidth]{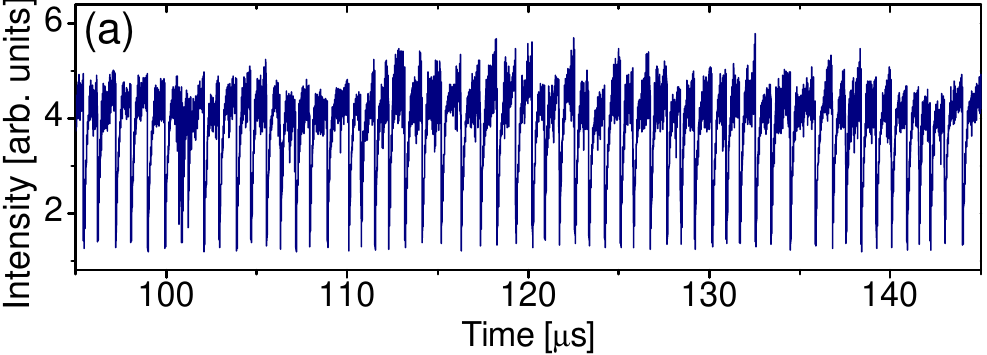}%[width=0.55\linewidth]
	\label{Fig2a}}
	\subfloat{
   	\includegraphics[width=0.275\linewidth]{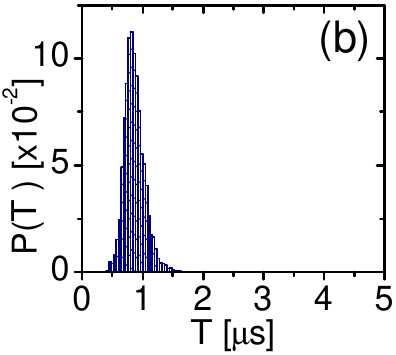}%[width=0.55\linewidth]
 	\label{Fig2b}}\\
	\subfloat{
   	\includegraphics[width=0.7\linewidth]{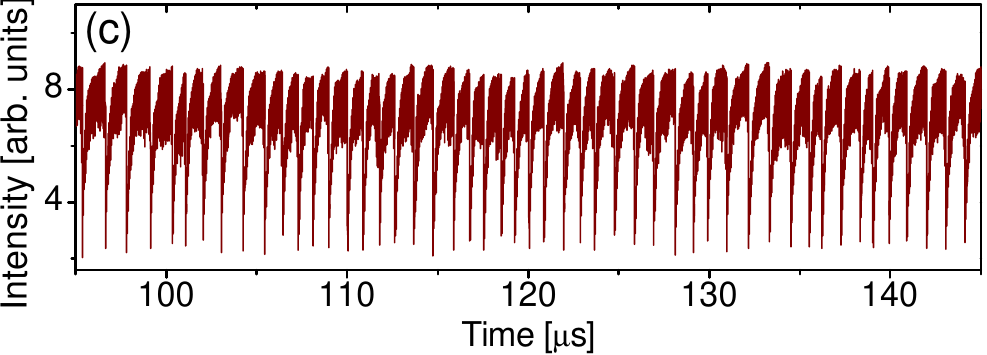}%[width=0.55\linewidth]
 	\label{Fig2c}}
	\subfloat{
   	\includegraphics[width=0.275\linewidth]{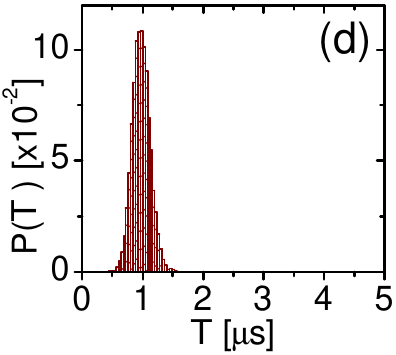}%[width=0.55\linewidth]
 	\label{Fig2d}}\\
	\subfloat{
   	\includegraphics[width=0.975\linewidth]{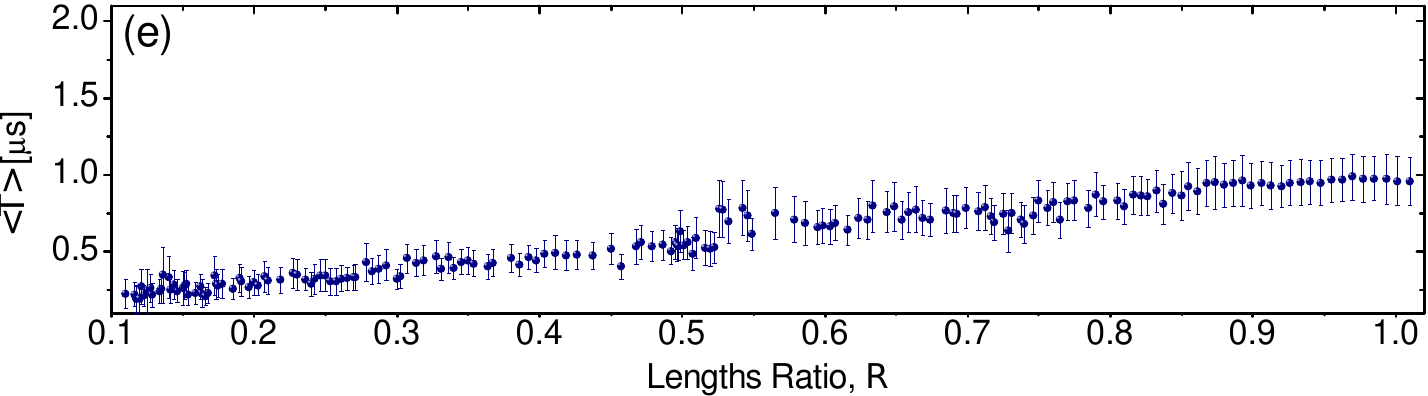}%[width=0.55\linewidth]
 	\label{Fig2e}}\\
	\subfloat{
   	\includegraphics[width=0.975\linewidth]{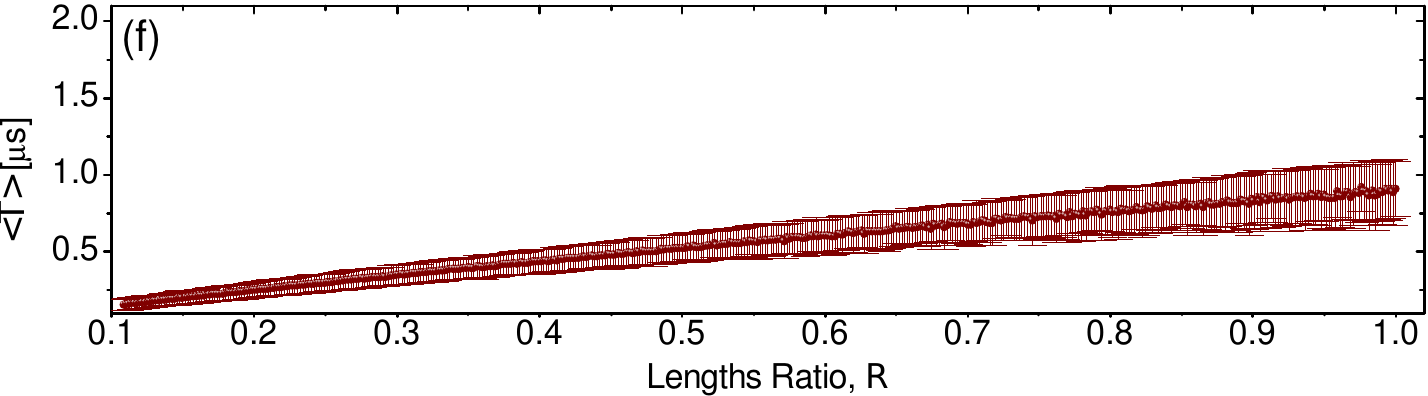}%[width=0.55\linewidth]
 	\label{Fig2f}}
\caption
{Single Feedback time series and chaos statistics. (a) Experimental segment of intensity time series for a single ($L1$ blocked) feedback configuration with $L_2 = 8.02$\,m.  (b) Histogram of the inter-spikes times of the time series presented in (a), showing an almost Gaussian shape with the long tail index $\alpha=1.8$. (c),(d) Numerical counterpart of (a) and (b), respectively. 
Calculations used the parameters showed in the methods section and produced long tail index $\alpha=2.0$. Both histograms are composed by approximately $10^4$ events. 
(e) Experimental average of inter-spike times $\langle T \rangle$ as function of the ratio length $R=L_2/L_1$ in a single feedback configuration maintaining $L_1$ blocked and scanning $L_2$ from 0.90 m to 8.02 m. (f) Numerical calculation equivalent to (e). The x-axis is normalized to 8.02m which is the value for the fixed arm $L_1$. In (e) and (f) the standard deviation $\sigma_T$ is represented by vertical bars. 
The colours (online) blue and red represent the experimental and the numerical calculations, respectively.} 
\label{Fig2}
\end{figure}

Chaotic spiking power drops in semiconductor laser with single feedback has been extensively studied \cite{sciamanna-nature2015}. 
We present its statistical properties here for the sake of comparison with the case of double feedback to be discussed later on.
A typical segment of an experimental 
series with a single feedback is shown in Fig.~\subref*{Fig2a}.  Power drops with a 
gaussian distribution is revealed in the corresponding histogram of Fig.~\subref*{Fig2b}.
Corresponding numerical-theoretical calculations are given in the time series segment of Fig.~\subref*{Fig2c}. 
The related histogram, in Fig.~\subref*{Fig2d}, also looks like a Gaussian probability 
distribution as in the experimental result.  

%%%%% FIGURE 2  %%%%%

From the histograms with more than $10^4$ drop events,  we  calculate the tail index {\bf $\alpha$} that represents the tail decay of the distributions \cite{McCulloch}. 
To remind,  pure Gaussian distributions have
$\alpha=2$ while the long tail Lorentz distribution has $\alpha=1$.  
The experimental probability distribution of Fig.~\subref*{Fig2b} presents 
a tail index $\alpha=1.8$, which is close to the Gaussian value. 
Numerical-theoretical results from Fig.~\subref*{Fig2d} for the histogram present a tail index $\alpha=2.0$. The similarity between the experiments and 
the theory is therefore verified in the statistics of spikes.
Scanning the $L_2$ cavity length with a single feedback over the range from $0.90$\,m to $8.02$\,m
gave us a large collection of time series from which we extracted the 
average $\langle T \rangle$ and standard deviation 
$\sigma_T$ of the inter-spike time as functions of the feedback time. 
Smooth, nearly constant, dependence for $\langle T \rangle$ was obtained collecting  $120$ equally separated runs moving $L_2$ 
from approximately $0.90$ m to $8.02$ m while $L_1$ was kept constant and equal to $8.02$ m during 
the entire scan as shown in Fig.~\subref*{Fig2e}.
The numerical solution of the rate equations  resulted with the same behavior and is given in Fig.~\subref*{Fig2f}.
The points in these graphics represent the mean time between consecutive LFF drops 
for time series of intensity with more than $10^{4}$ events each. The vertical 
bars represent the standard deviation. 
In both experimental and numerical data the $\alpha$ index  was always near $1.9$, confirming 
Gaussian shape for the histograms, independent of 
feedback time in the studied range. Small feedback times ($L_2 < 0.50$\,m) are known to make 
the system depart from LFF dynamics \cite{heilShortCavity}.
The parameters of the equations were taken to produce the observed quantitative coincidence 
with the experimental values and are shown in the methods section. 
These results for a single feedback dynamics were taken for comparison with the striking new results from  
the double feedback case, to be presented next.

\subsection{Chaos with Double Feedback}

\begin{figure}[!h]
\centering
	\subfloat{
	\includegraphics[width=0.975\linewidth]{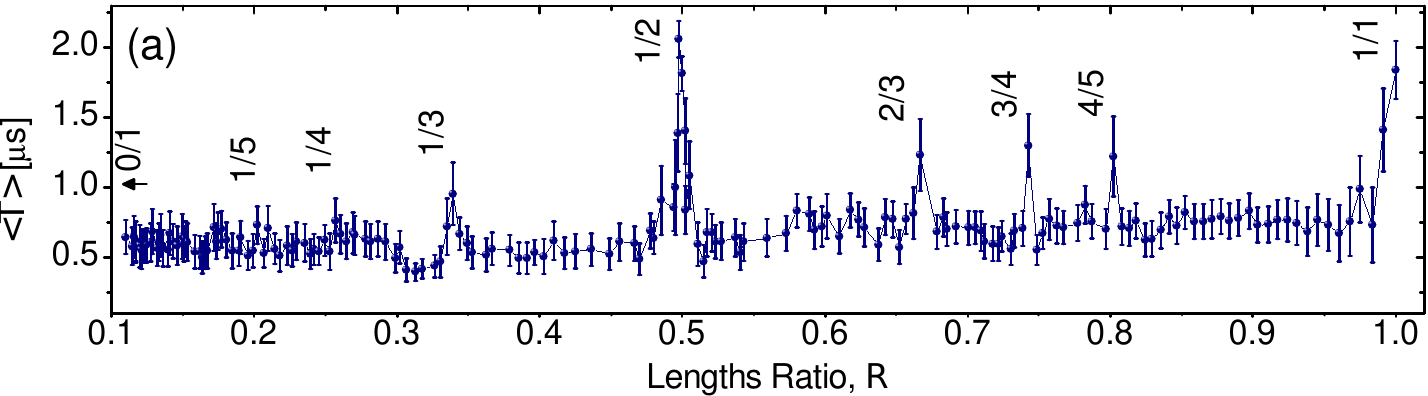}%[width=0.55\linewidth]
 	\label{Fig3a}}\\
	\subfloat{
   	\includegraphics[width=0.975\linewidth]{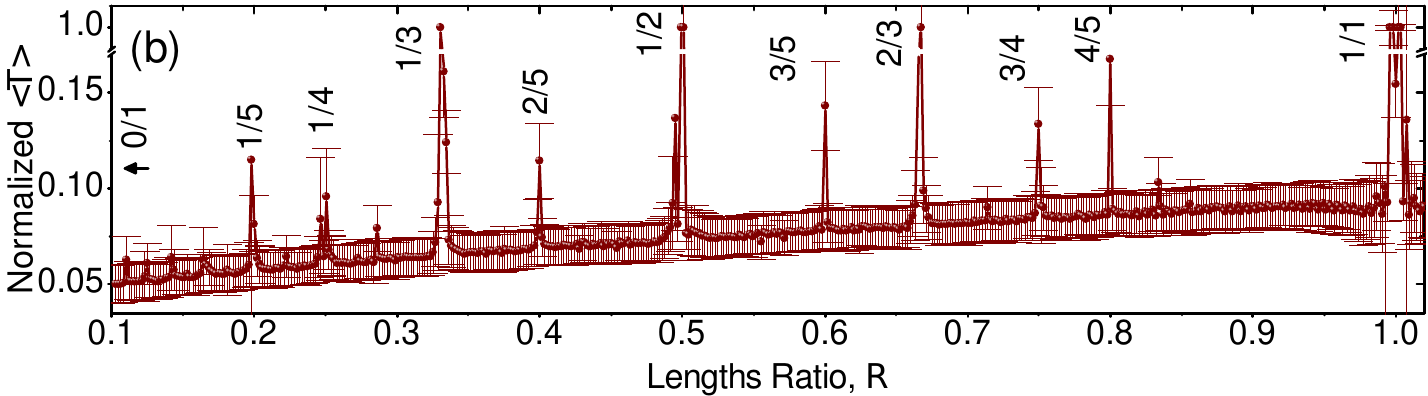}%[width=0.55\linewidth]
 	\label{Fig3b}}\\
	\subfloat{
   	\includegraphics[width=0.7\linewidth]{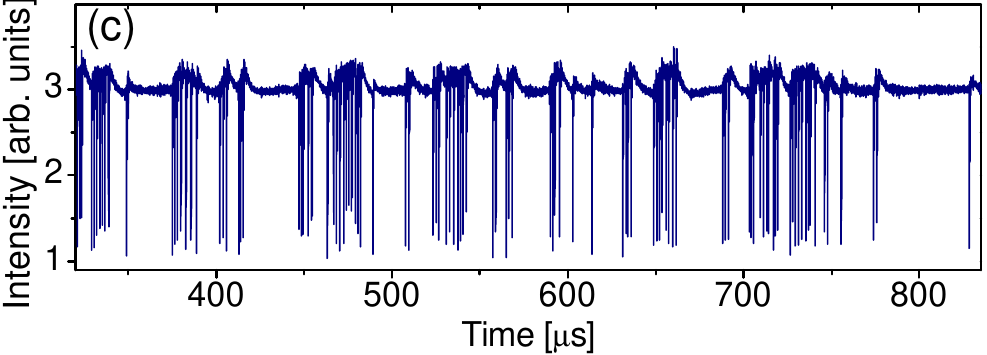}%[width=0.55\linewidth]
 	\label{Fig3c}}
	\subfloat{
   	\includegraphics[width=0.275\linewidth]{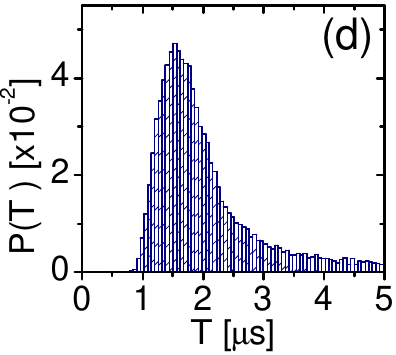}%[width=0.55\linewidth]
 	\label{Fig3d}}\\
	\subfloat{
   	\includegraphics[width=0.7\linewidth]{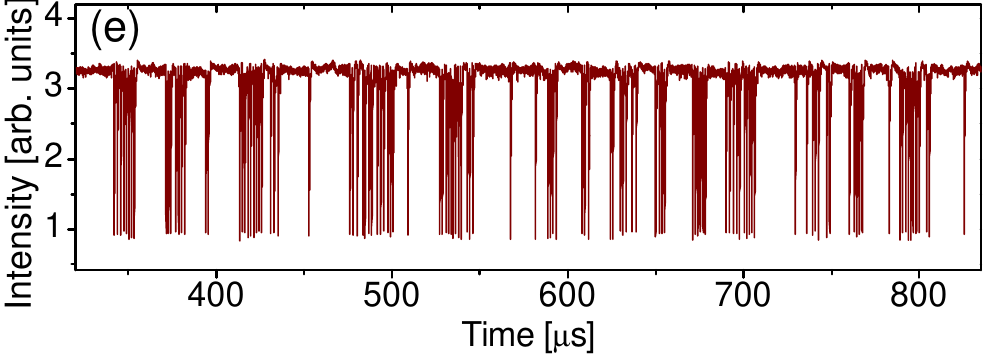}%[width=0.55\linewidth]
 	\label{Fig3e}}
	\subfloat{
   	\includegraphics[width=0.275\linewidth]{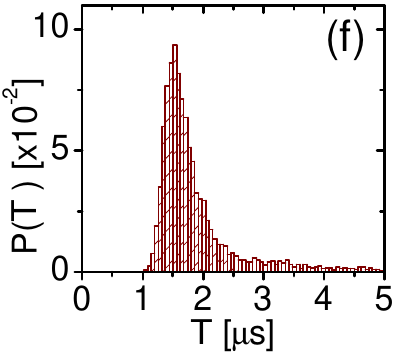}%[width=0.55\linewidth]
 	\label{Fig3f}}
\caption{Double Feedback time series and chaos statistics. (a) Experimental average of inter-spike times $\langle T \rangle$ dependance on $R=L_2/L_1$ maintaining $L_1=8.02$ m constant and scanning $L_2$ from 0.90 m to 8.02 m. (b) Numerical calculation equivalent to (a) with the y-axis is normalized to the time series duration.
(c) Experimental segment of intensity time series for a double feedback configuration with $R=0.498$ and $L_1 = 8.02$ m.  (d) The histogram of the inter-spikes times corresponding to (c), showing a long-tail shape with the tail index $\alpha=1.1$. (e), (f) Numerical calculation equivalent to (c) and (d) respectively for $R=0.4946$ and using the parameters showed in the methods section. The theoretical tail index obtained is also $\alpha=1.1$. Both histograms are composed by approximately $10^4$ events. The colours (online) blue and red represent the experimental and the numerical calculations, respectively.
}
\label{Fig3}
\end{figure}

The inclusion of the second feedback in the laser changes drastically its dynamics as compared to what was reported  experimental and theoretically for one delay. 
For the double feedback experiments the fixed length arm $L_1$, with feedback time $\tau_1$, is unblocked and give the same (as good as possible) amount of field return 
established for the variable arm $L_2$.
Repeating the $120$ experimental steps for the values of $\tau_2$, like was done in the one feedback case presented in Fig.~\subref*{Fig2e}, completely different scenarios 
are found.
This is shown in Fig.~\subref*{Fig3a} which gives the measured dependence of the average time $\langle T \rangle$ as function of the feedback lengths ratio $R$. The length increments $\Delta L_2$ correspond to  $\approx$ 6 cm. 
The theoretical counterpart, calculated with steps of 1.5 cm, is shown in Fig.~\subref*{Fig3b}. 
%
%%%% FIGURE 3 - TYPICAL SERIES, HISTO AND R SCAN - DOUBLE FB %%%%%
%
One clearly see the sensitivity of $\langle T \rangle$ on this parameter.
When $R$ is close to rational numbers with small denominators, the  
low frequency fluctuation drops of intensity become less frequent due to the appearance of long 
intervals without drops.
As a consequence, the mean time between drops $\langle T \rangle$ increases and the data 
variance diverges. Notice that in both, experimental and numerical, the effect observed occurs with external cavity length steps $\approx$ $\times 10^4$ optical wavelengths. 
Farey sequences, defined as the set of rational numbers written as the ratio $p/q$ of two natural numbers,
are known to manifest in frequency locking or entrainment between oscillators. 
These physical phenomena deal mostly with periodic, quasiperiodic motions and the onset of chaos  \cite{Xu-PRA}. 
In our results the Farey hierarchy plays a critical role in changing chaos statistics.   
Bunching of 
spikes appear around the small denominator ratio, also called low order Farey numbers, in a very sensitive way with respect to the ratio $R$. 
Those transitions from statistic behavior with Gaussian shape to spiking burst long-tail distributions are characterized by peaks of $\langle T \rangle$ near Farey fractions during the scan of $R$. The most significant peak in value of $\langle T \rangle$ happen around $R=1/2$. Outside those Farey ratios we have LFF chaos with $\langle T \rangle$ and $\sigma_T$ quite insensitive to the changes in $R$. This indicates Gaussian profiles for the inter-spikes time histogram.  As $R$ increase towards $1/2$ (similar to what happens near 1/1 and higher order Farey ratios) $\langle T \rangle$ and the respective variance increase drastically indicating transitions to bursting regime of spikes. Crossing the value $R=1/2$ these statistics parameters decrease again returning to the Gaussian chaos. As will be shown latter on, details of those transitions, including stationary regime and chaos without LFF drops, also occur for values of $R$ very close to these special ratios.   
Figure \subref*{Fig3c}   
presents an interval of the experimental time series in the case of spiking burst behavior 
for a length ratio $R=L_2/L_1\approx0.498$, i. e., near $1/2$. 
The corresponding histogram of the $10^4$ inter-spike times is shown in Fig.~\subref*{Fig3d}. 
Unlike the Gaussian distribution presented in the single feedback, and also in the double feedback case with $R$ far from low order Farey numbers (1/2, 1/3, 2/3, 1/1, etc), in this case a heavy tail distribution is observed. For this experimental data the calculated long tail index is $\alpha=1.1$, 
confirming the presence of an expressive long tail in the probability distribution $P(T)$ due to the long time interval between bunches of spikes. Figs. \subref*{Fig3e} and \subref*{Fig3f} 
present the numerical calculation for $R=0.4946$ showing the same non-Gaussian distribution and spiking burst behavior as in the 
experimental results. The associated histogram, also composed by approximately $10^4$ intervals, also has a long tail index $\alpha=1.1$.

Definitively the model equations fit the transitions of 
chaos statistics observed. Better quantitative comparison would ask for further exploration of the equations parameters.
In contrast to the double feedback results, experiments and calculations with single feedback (with our selected parameters) never show spiking bursts throughout 
the whole range of variation of $L_2$ as indirectly indicated in Figs. \subref*{Fig2e} and \subref*{Fig2f}. It is important to reinforce that, different from other previously published works which include 
stochastic terms in the models, our equations are fully deterministic and so the theoretical results depend only on the system parameters.

%%%%% SECTION: DISCUSSION %%%%
%
\section{Discussion}

\subsection{Stationary Solutions and Bifurcations}

The stationary solutions ($E_s$, $N_s$, $\omega_s$) for the Eqs.~(\ref{eq:LK2FB-E})-(\ref{eq:LK2FB-N}) are obtained setting $\frac{dE(t)}{dt}=0$, $\frac{d\Phi(t)}{dt}=(\omega_s - \omega_0)$, $\frac{dN(t)}{dt}=0$ and using the condition $E(t)=E(t-\tau_1)=E(t-\tau_2)=E_s$. While $\omega_0$, the solitary laser frequency, is always constant, $\omega_s$, the frequency of the compound system including the two cavities, vary in time and will be constant for the stationary state. In general, multiple constant values for $\omega_s$ result from the steady state condition. In each solution, the dynamical variable $\Phi(t)$ reads $\Phi(t)=(\omega_s-\omega_0)t$. A simplification of the equations is possible setting the gain saturation coefficient equal to zero ($\epsilon=0$) as the general features of the system are still preserved. Equations (\ref{eq:LK2FB-E})-(\ref{eq:LK2FB-N}), for the stationary solution, then became: 
\begin{eqnarray}
 \omega_s - \omega_0 =  -\kappa_1[\overline{\alpha}  \cos{(\omega_s \tau_1)} + \sin{(\omega_s \tau_1)}] \nonumber \\ 
 - \kappa_2[\overline{\alpha}  \cos{(\omega_s \tau_2)} + \sin{(\omega_s \tau_2)}] \label{eq:StationarySolutionsLK2FB-omega} 
 \end{eqnarray}
 \begin{eqnarray}
N_s&=&N_0+\frac{\gamma_p}{G_0}-\frac{2}{G_0}[\kappa_1\cos(\omega_s\tau_1)+\kappa_2\cos(\omega_s\tau_2)] \label{eq:StationarySolutionsLK2FB-N}\\ 
E_s&=&\frac{J-N_s/\tau_s}{\gamma_p-2[\kappa_1\cos(\omega_s\tau_1)+\kappa_2\cos(\omega_s\tau_2)]}.\label{eq:StationarySolutionsLK2FB-E}
\end{eqnarray}\\
Equation (\ref{eq:StationarySolutionsLK2FB-omega}) is transcendental and has no analytical solution. Therefore, numerical procedure is necessary to obtain $\omega_s$ for a given set of parameters. All values obtained for $\omega_s$ are then introduced in Eqs. (\ref{eq:StationarySolutionsLK2FB-N}) and (\ref{eq:StationarySolutionsLK2FB-E}) producing the steady state values of $N_s$ and $E_s$. The graphical inspection of the $\omega_s$ solutions can be done defining $F(\omega_s)$ and $G(\omega_s)$ as the left-hand and right-hand sides of Eq.~(\ref{eq:StationarySolutionsLK2FB-omega}), respectively.

\begin{figure}[!h]
\centering
	\subfloat{
	\includegraphics[width=0.48\linewidth]{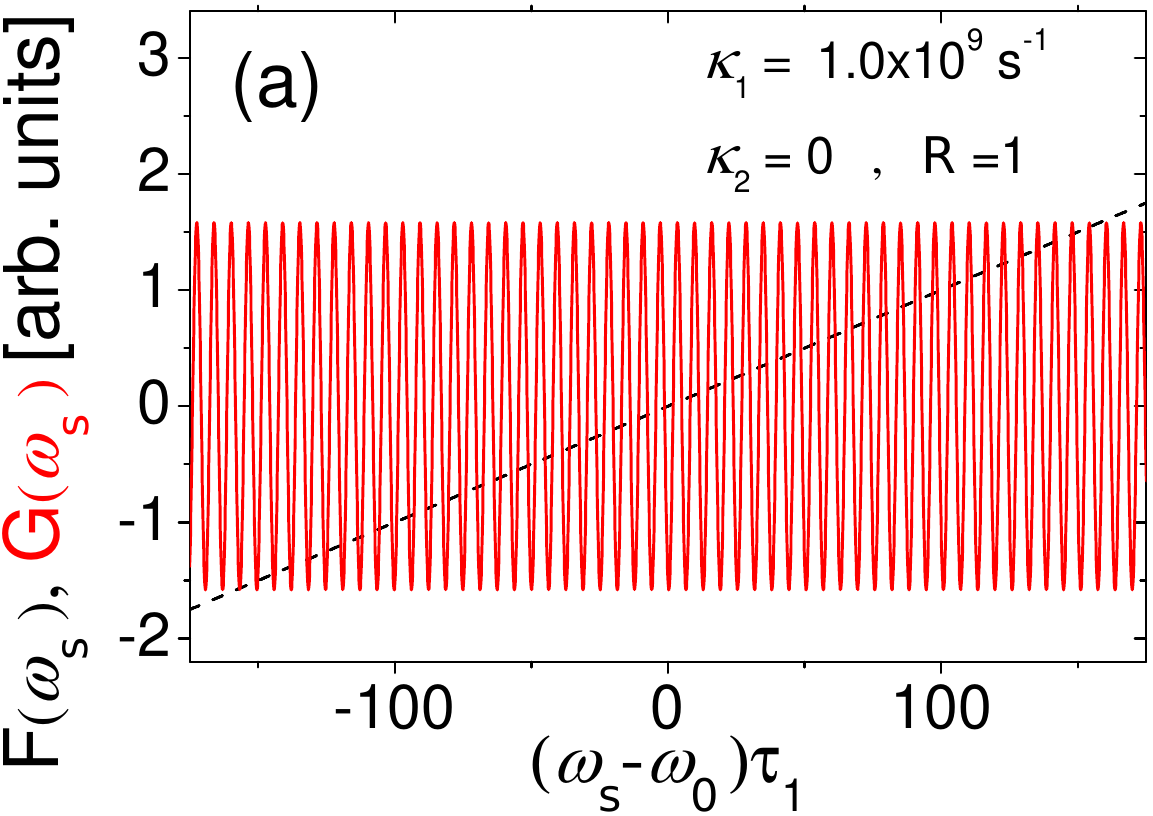}%[width=0.55\linewidth]
 	\label{Fig4a}}
	\subfloat{
	\includegraphics[width=0.48\linewidth]{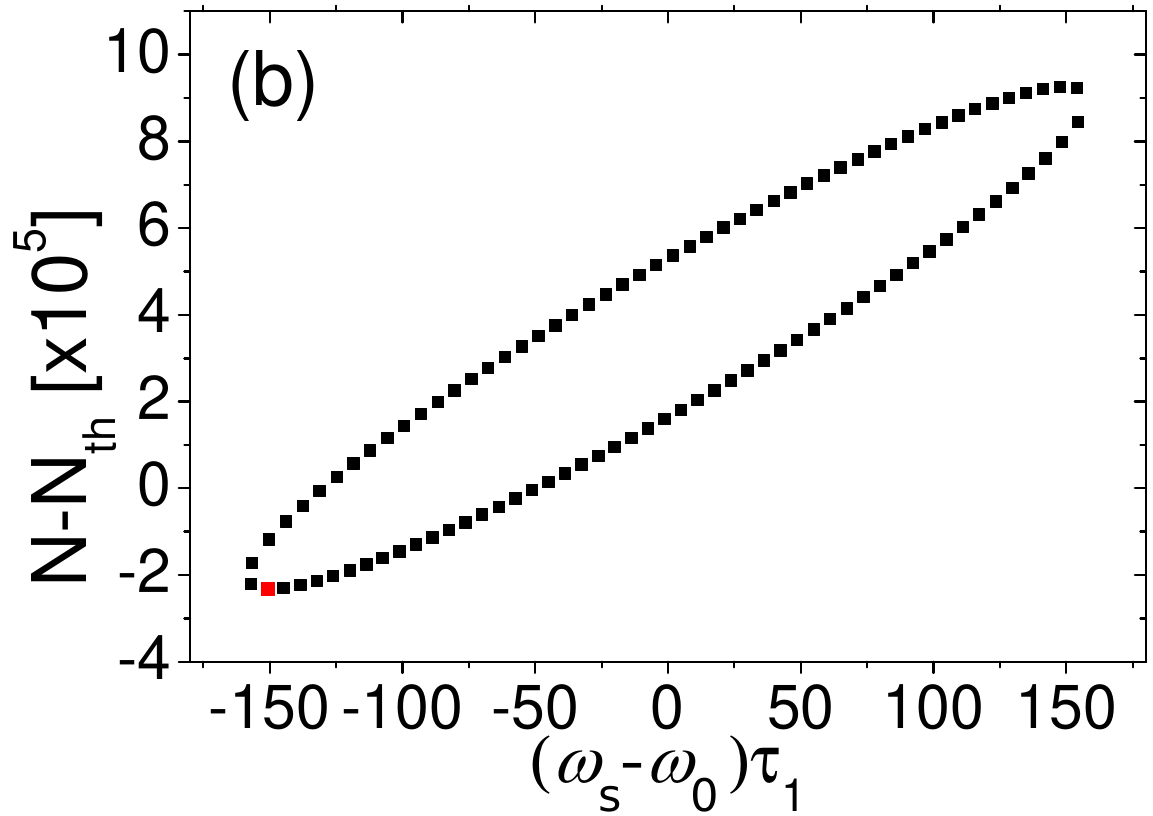}%[width=0.55\linewidth]
 	\label{Fig4b}}\\
	\subfloat{
	\includegraphics[width=0.48\linewidth]{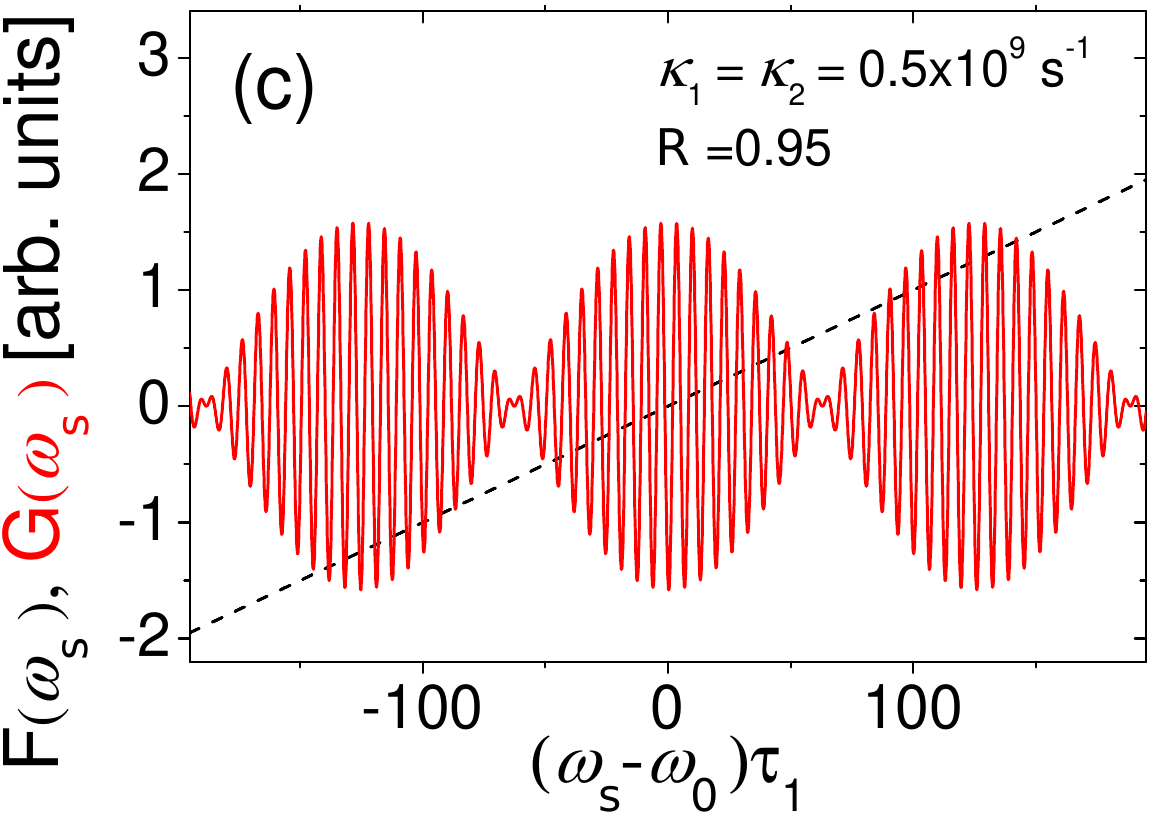}%[width=0.55\linewidth]
 	\label{Fig4c}}
	\subfloat{
	\includegraphics[width=0.48\linewidth]{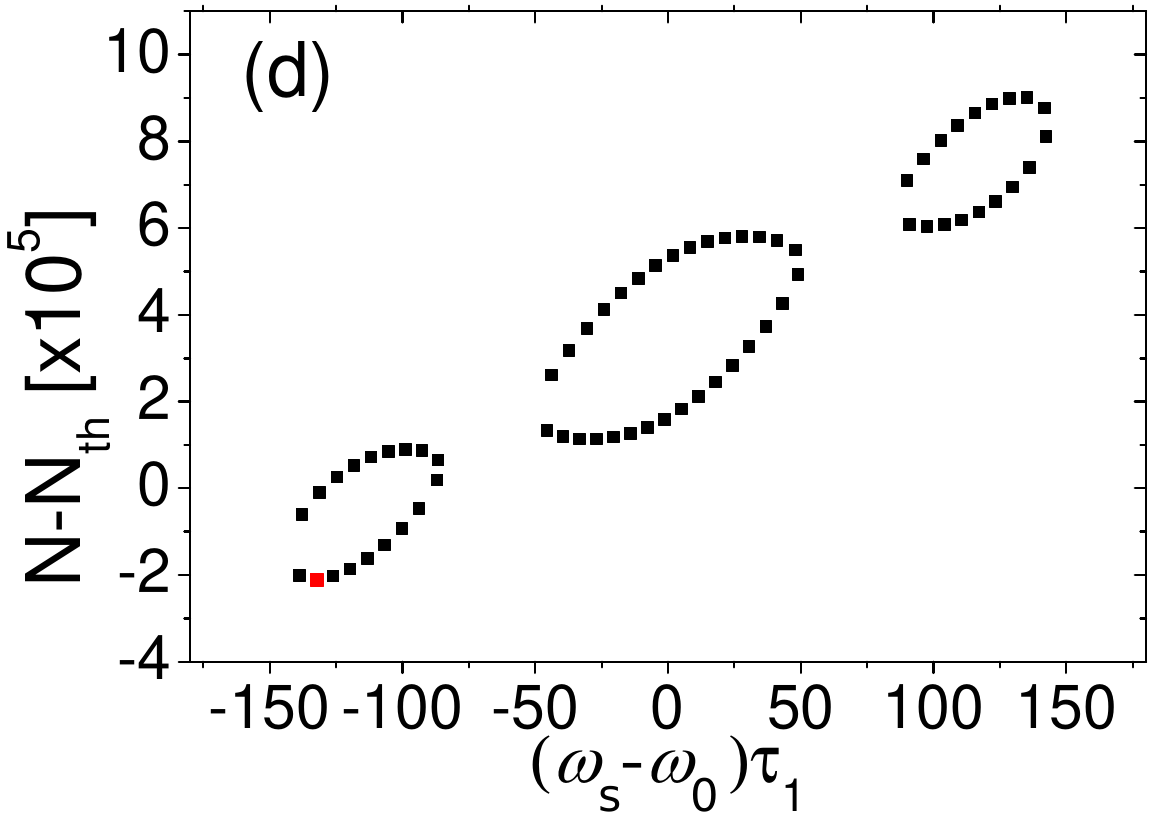}%[width=0.55\linewidth]
 	\label{Fig4d}}
\caption{Stationary solutions. (a) Single feedback: The crossing of $F(\omega_s)$ (dashed black line) with $G(\omega_s)$ (red line) for moderate feedback strength ($\kappa_1=1.0\times10^9$\,s$^{-1}$, $\kappa_2=0$) showing tens of crossing points creating pairs of stable-unstable fixed points (solutions $\omega_s$). (b) Representation of the solutions $\omega_s$ from (a) in the ($\Delta \omega \tau_1$,\,$\Delta N$) plane. The ellipse that supports the fixed points is clearly visible. The lower point in the ellipse, represented by a red point (online), is called the Maximum Gain Mode (MGM). It is the mode with the lower threshold of operation. (c) Double feedback: The same of (a) but with the two external cavities sending light back to the laser with equal strength ($\kappa_1=\kappa_2=\kappa=0.5\times10^9$\,s$^{-1}$). (d) For double feedback, those solutions occur over three ellipses due to the beating effect showed in (c). It changes the form of the attractor skeleton when represented in the phase space projection ($\Delta \omega \tau_1$,\,$\Delta N$). 
}
\label{Fig4}
\end{figure}

For the single feedback case, i.e. $L_1$ blocked (numerically we set $\kappa_1=0$), the stationary solutions are shown in Figs. \subref*{Fig4a} and \subref*{Fig4b} for moderate feedback strength ($\kappa_2=1.0\times10^9$\,s$^{-1}$) and for $R=1$ which means $L_2$ is equal to the length of the fixed arm $L_1$. For this case, the graphical solution of Eq.~(\ref{eq:StationarySolutionsLK2FB-omega}) is presented in Fig.~\subref*{Fig4a} where $F(\omega_s)$ (dashed black line) and $G(\omega_s)$ (red line) are showed in terms of $\omega_s$ (multiplied by the roundtrip time of the fixed external cavity $\tau_1$, here blocked). 
The crossing points means that $F(\omega_s)=G(\omega_s)$. It gives the frequencies solutions $\omega_s$ of Eq.~(\ref{eq:StationarySolutionsLK2FB-omega}). The complete stationary solutions $(E_s,N_s,\omega_s)$ are obtained using the Eqs. (\ref{eq:StationarySolutionsLK2FB-N}) and (\ref{eq:StationarySolutionsLK2FB-E}). 
These solutions can be represented in a projection of the phase space ($\Delta \omega \tau_1$,$\Delta N$) as it is shown in Fig.~\subref*{Fig4b}, where $\Delta \omega = \omega_s - \omega_0$ and $\Delta N  = N_s-N_{th,sol}$. They appear over an ellipse \cite{sano} which analytical equation can be obtained by manipulating Eqs. (\ref{eq:StationarySolutionsLK2FB-omega}) and (\ref{eq:StationarySolutionsLK2FB-N}). 
Here, the steady solutions were obtained numerically. The term $N_{th,sol}=N_0-\gamma_p/G_0$ is the threshold value for $N$ when the laser is being operated in the solitary mode (with no feedback). 

The appearance of the solutions, by changing parameters like the feedback strength or the time delay, happen through a saddle-node bifurcation of cycles when $F(\omega_s)$ touch $G(\omega_s)$. The result of this crossing process is the creation of two new values for $\omega_s$ in each bifurcation. These solutions are classified as limit cycles, one stable mode (lower points in the ellipse) and the equivalent unstable antimode (upper points in the ellipse). The comb of stable resonant frequencies $\omega_s$ provided by the compound cavity are called external cavities modes (ECMs). The center of the ellipse represents the solitary laser mode when the laser has no external feedback ($\kappa_1=\kappa_2=0$). The mode with the largest gain, is called Maximum Gain Mode (MGM). In Fig.~\subref*{Fig4b} it is represented by a red square. It is the lower point of the ellipse in the ($\Delta \omega$,\,$\Delta N$) phase space projection representation. This mode should operate with the lower value of $N$ and the higher value of $E$ (and consequently higher intensity $|E|^2$). It is the mode with the lowest pump current threshold $J_{th,MGM}$. The stability of this mode in numerical calculations is highly sensitive with respect to the lasers parameters \cite{masoller3,jhonPRL2008}.

When no mode on the ellipse can stabilizes the laser, the LFF regime in which the emitted intensity of the laser $|E|^2$ presents irregular spike drops is observed \cite{masoller3,jhonPRL2008,sacher, sano, hohl,  giudici, sukow, mulet, eguia1}. 
A well accepted theoretical explanation of this chaotic behaviour was given by Sano \cite{sano}. Accordingly, the ECMs are destabilized and became quasi-attractors causing mode hopping between them through wave mixing process.  
According to Sano, the trajectory of the system in the phase space projection surrounds the quasi-attractors (fixed points represented in the ellipses) in a local chaos transient stepping to the next ECM with lower frequency (to the left in the ellipse). The mode hopping repeat until the system reach the neighborhood of the MGM.  
The system remains surrounding the MGM until the trajectory crosses the stable manifold of the saddle anti-mode being repulsed from it and making the system experiment a frequency shift backing near to the solitary frequency operation. In this process the variable $N$ increases approaching to its threshold value $N_{th}$ and the intensity $|E|^2$ drops in a very short time scale. This entire process repeat in an irregular way characterizing the chaotic LFF regime. 

%%%%% FIGURE 4  %%%%%

For the double feedback configuration, the right-hand side $G(\omega_s)$ of Eq.~(\ref{eq:StationarySolutionsLK2FB-omega}) has two contributions giving a beating effect  due to the difference $\Delta \tau = \tau_1 - \tau_2$. This is shown in Fig.~\subref*{Fig4c} for $R=0.95$ and $\kappa=0.5\times10^9$\,s$^{-1}$. 
In the calculations we used $\kappa_1=\kappa_2=\kappa$, which means that both arms of optical feedback are sending light back to the gain medium with the same strength. Experimentally, to provide that, equal laser threshold reduction due to each optical feedback alone  was monitored and adjusted through mirrors alignment.
It is important to notice that this pattern is not a result of optical frequency interferometry, once $\Delta \tau$ is very large if compared to the optical period as in our experimental and theoretical scan of $R$. Figure \subref*{Fig4d} shows that in this case, due to this beating effect between the external cavities lengths, the attractor skeleton represented by the points in the plane ($\Delta \omega \tau_1$,$\Delta N$) now settle over three ellipses. The scan of $R$ can also produce five, seven, nine and more ellipses. 
These multiple ellipses attractors result from the beating modulation, presented in Fig.~\subref*{Fig4c}. This effect is directly related to the trajectory in phase space \cite{sano} changing the chaotic dynamics of the systems variables incorporating bursting in the LFF regime.

\subsection{Laser Threshold and Chaos Statistics}

A clarifying view of the system physical behavior with double feedback can be reach examining the threshold $J_{th,MGM}$ variation of the MGM as lengths ratio $R$ is varied. The MGM is the most relevant fixed point of the system. Changes in it, through parameters variation, may induce bifurcations in the system. 
Even in chaotic regime of operation, the MGM remains the fixed point around which the trajectory spends most of the time. 
Actually, due to the infinity dimension of the system, the size of MGM basin of attraction must be small enough that usual trajectories never penetrate it. This is consistent with our deterministic numerical results and experiments. 
Futhermore, even if the basin of attraction is visited \cite{masoller3}, the injection of external optical signal or noise can excite LFF drops. This justifies the proposition that this system is excitable \cite{giudici}. An alternative view point is to argue that the ultrafast deterministic oscillations in the dynamics excite the drops in an effective slow time scale attractor \cite{jhonPRL2008}.

\begin{figure}[!h]
\centering
	\subfloat{
	\includegraphics[width=0.975\linewidth]{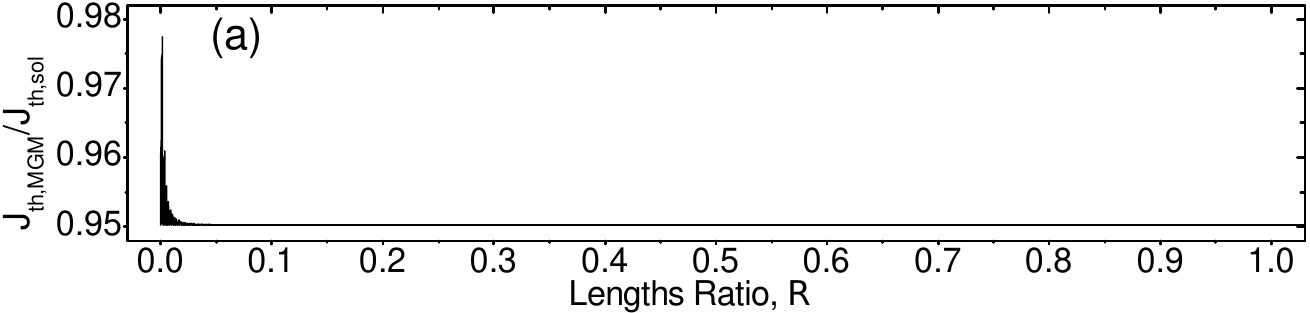}%[width=0.55\linewidth]
 	\label{Fig5a}}\\
	\subfloat{
   	\includegraphics[width=0.975\linewidth]{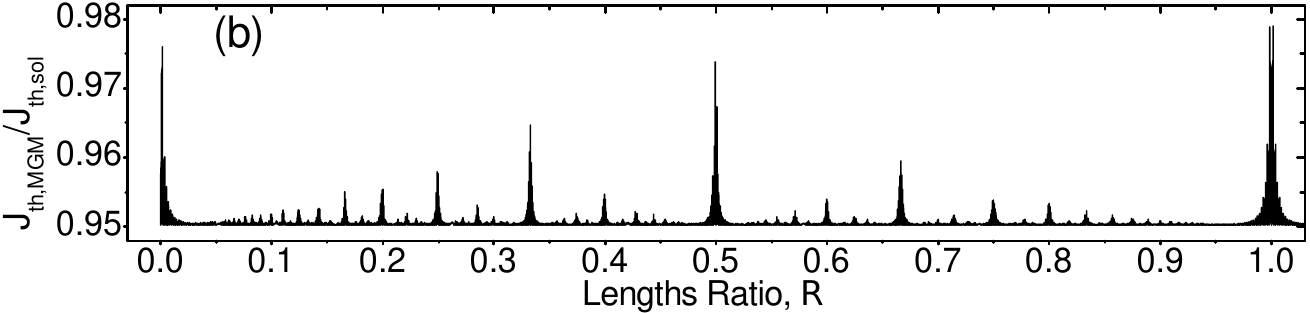}%[width=0.55\linewidth]
 	\label{Fig5b}}\\
	\subfloat{
   	\includegraphics[width=0.975\linewidth]{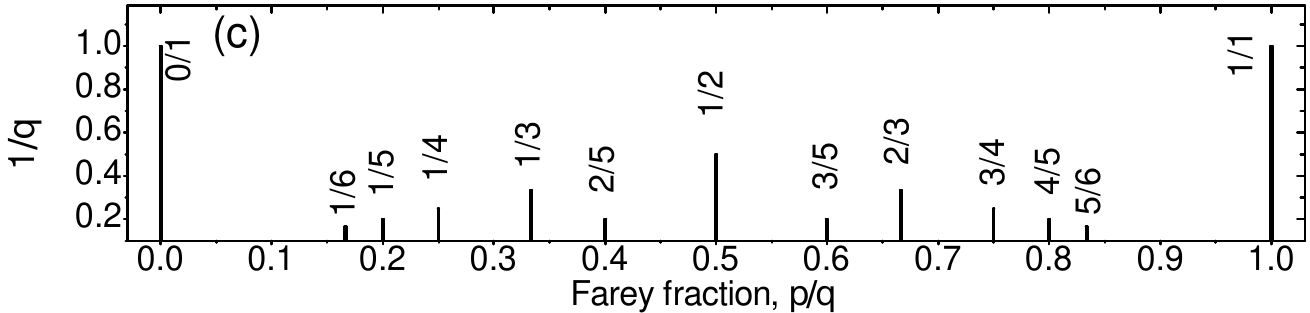}%[width=0.55\linewidth]
 	\label{Fig5c}}
\caption{Analytical MGM current threshold $J_{th,MGM}$ dependence on $R$ and Farey fractions. MGM current threshold $J_{th,MGM}$ variation with single feedback (a) and with double feedback (b). The vertical axes are normalized to the solitary current threshold $J_{th,sol}$. (c) Order sixth Farey sequence (Farey numbers with denominators $q\leq6$) with vertical axis representing the Farey number order as $1/q$.}
\label{Fig5}
\end{figure}

Differently from the single feedback case, where the fixed points ($E_s$, $N_s$, $\omega_s$) when plotted in the phase space projection ($\omega$, $N$) remain over a single ellipse \cite{sano}, the double feedback makes these points to be distributed in more than one ellipse changing the attractor in such a way that trajectories reach certain values producing non-gaussian - long tail - time distribution. 
The MGM threshold current $J_{th,MGM}$ dependence on $R$ reveals the effect of the parameters scan on the system dynamics changes.
Figures \subref*{Fig5a} and \subref*{Fig5b} show how the system change its MGM threshold when it is subjected to a single and a double optical feedback, respectively. It shows analytically how $J_{th,MGM}$ (normalized to the solitary threshold current $J_{th,sol}$) depends on the scan of $R$, which for the single cavity means the variation of the corresponding feedback time. 
%
%%%% FIGURE 5 - SCAN Jth AND FAREY - ANALITICAL %%%%%

For the laser to oscillate, with some non-null emission, the pump current $J$ has to be above $J_{th,MGM}$. It can be seen in Fig.~\subref*{Fig5a} that the threshold current is reduced with respect to the solitary laser ($J_{th,MGM}<J_{th,sol} $ in about 5\%).   
Notice that $J_{th,MGM}$ is constant for most values of $R$ except near $R=0$. Near this value, Figs. \subref*{Fig5a} and \subref*{Fig5b} are similar.
To be reminded, for single feedback (Fig.~\subref*{Fig5a}) the $R$ near zero means that the only active external cavity becomes comparable to the inner semiconductor laser cavity. 
It gives feedback time sufficient short to compete with the relaxations dynamics in the laser. In fact this range of values do not contemplate the occurrence of LFF \cite{fisher} and it was not observed in our experiments.

Double feedback, as shown in Fig.~\subref*{Fig5b}, gives a structure to the MGM threshold current as function of $R$. Its reduction has a base line equal to the single feedback case apart from the regions where $R$ is near a low order Farey number.  
The changes in $J_{th,MGM}$ is more significant with peaks according to the Farey order $p/q$ in $R$. 
Throughout the range $0<R<1$ equal Farey denominators $q$ give similar peaks for the threshold reduction. 
For the purpose of comparison with Fig.~\subref*{Fig5b}, the sixth order Farey sequence (all the fractions p/q with $q\leq6$) is shown in Fig.~\subref*{Fig5c}. The vertical axis of Fig.~\subref*{Fig5c} represent the Farey order as $1/q$. It is remarkable the comparison of these orders with peaks sizes of $J_{th,MGM}$  from Fig.~\subref*{Fig5b}. The lower the Farey number order, the stronger the effect on the threshold reduction.
Previous work related to our report remotes to the first studies of instabilities in light propagation through ring optical cavities.  
Ikeda and Mizuno have studied instabilities in double cavity nonlinear optical resonators \cite{IkedaPRL1985,ikedaIEEE1985,mizuno}. 
They argue with "frustrated instabilities" in a non-chaotic system with many oscillation modes. 
In their system, the frustration behavior always happens around the Farey fractions as in ours. 
However, as we treat an active laser system, our study is different, referring to chaotic dynamics and statistics.

\begin{figure}[!h]
\centering
	\subfloat{
	\includegraphics[width=0.975\linewidth]{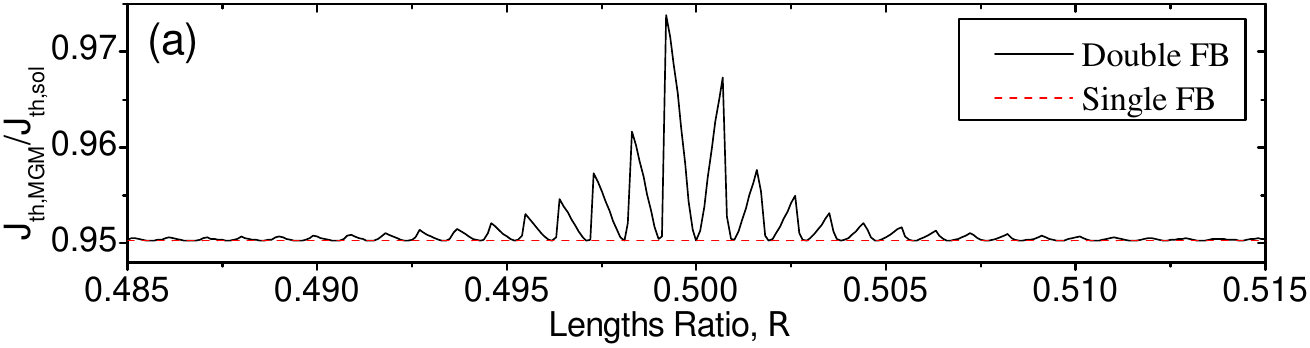}%[width=0.55\linewidth]
 	\label{Fig6a}}\\
	\subfloat{
   	\includegraphics[width=0.975\linewidth]{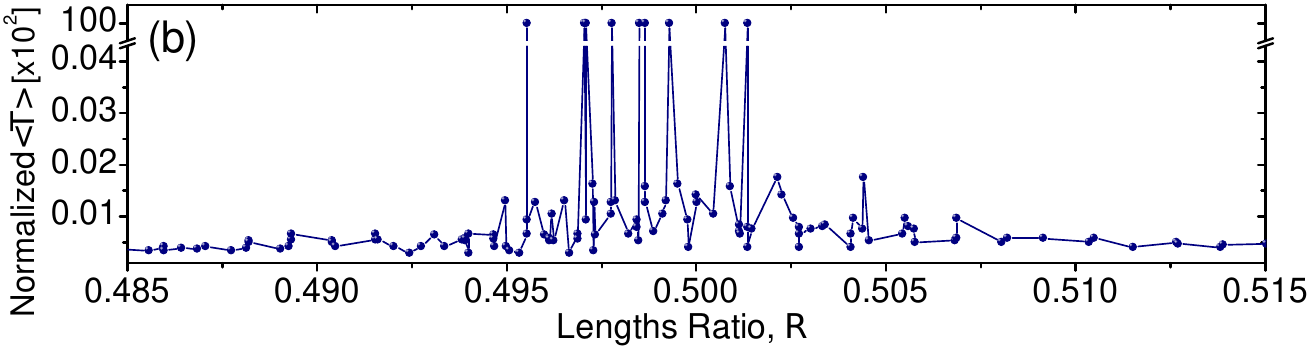}%[width=0.55\linewidth]
 	\label{Fig6b}}\\
	\subfloat{
   	\includegraphics[width=0.975\linewidth]{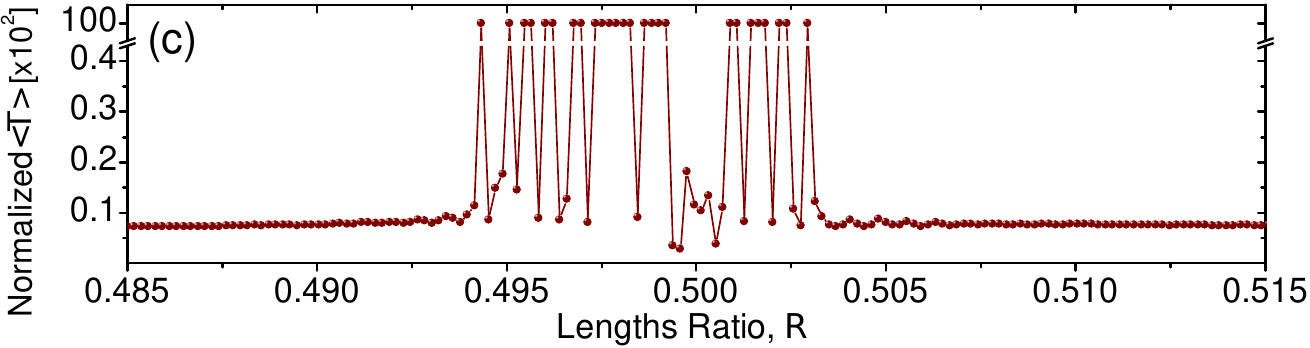}%[width=0.55\linewidth]
 	\label{Fig6c}}
\caption{MGM current threshold and inter-spike time averaged near the Farey fraction $R=1/2$. (a) Analytical dependence of $J_{th,MGM}$ on $R$ presenting its fluctuations for the single feedback (red dashed line online) and the double feedback (black line online) cases around $R=1/2$. (b) Experimental  $\langle T \rangle$ near $R=1/2$ showing its strong fluctuations through the scan of $R$. (c) Numerical counter part of (b). In both cases y-axis is normalized to the time series duration. Considering the relevant orders of magnitudes variations of $\langle T \rangle$ no error bars or variance are indicated (b) and (c). }
\label{Fig6}
\end{figure}

Concerning the significant regions where the system shows spike-bursts we focus on $R$ values zooming the calculations close to Farey numbers.
Fine details in $J_{th,MGM}$ indicate oscillations, as shown in Fig.~\subref*{Fig6a}. Those behavior do not correspond to optical interferometric oscillations. They have to do with the difference between the two feedback time delays as discussed in the previous sections.  Such oscillations have an impact in the chaotic statistics. 
Experimental and numerically calculated mean inter-spike times $\langle T \rangle$ through the same range of Fig.~\subref*{Fig6a} are shown in Figs. \subref*{Fig6b} and \subref*{Fig6c}, respectively.  
%
%
%%%% FIGURE 6 - SCAN NEAR 1/2 - EXP AND TEO - ZOOM SCAN Jth %%%%%

Approaching the $1/2$ quotient, the system maintains the mean $\langle T \rangle$ and its variance with an almost constant value until $R \approx 0.494$ while the laser operates in the LFF regime with Gaussian statistics of spikes. 
Proceeding towards $1/2$ makes $\langle T \rangle$ to grow and the variance diverges. The laser then exhibits bursting operation, in which we observe very long intervals without LFF as was shown in Figs. \subref*{Fig3c} and \subref*{Fig3e}. 
Varying $R$ further, $\langle T \rangle$ becomes larger as the quiescent time between bursts increases. It may even reach the condition where the intensity time series presents no LFF drops. 
Apart from LFF dynamics and statistics, the laser goes through many different forms of oscillation including local chaos around the MGM, periodic oscillation and continuous wave operation. None of these regimes is inspect in our study.
To obtain $\langle T \rangle$ we have used a LFF pulse counting procedure and therefore, any information about the real dynamical state of the intervals without LFF drops could not be accessed through our procedure.
When no LFF pulse was found in an entire time series, we placed $\langle T \rangle$ as the total duration of the series as shown in the top values of $\langle T \rangle$ in Figs. \subref*{Fig6b} and \subref*{Fig6c}. 
Continuous increasing of $R$ induces large variations with ups and downs of $\langle T \rangle$ similar to what happens with threshold current $J_{th,MGM}$ for the MGM. This indicates that the succession of transitions between chaotic LFF regime with Gaussian statistics to non-LFF regime are physically produced by the variations of the compound laser threshold in our model. 
Increasing $R$ beyond $1/2$ shows again the regions of different instabilities until chaotic LFF regime with Gaussian statistics is recovered. Experimentally and numerically, no exact symmetry is observed with respect to $R=1/2$. This is consistent with the analytical result for $J_{th,MGM}$ in Fig.~\subref*{Fig6a}.    
It is also observed that on the exact value of the rational Farey fraction $R=1/2$, $J_{th,MGM}$ is equal to the single feedback threshold value (red dashed line). For this value, Figs. \subref*{Fig6b} and \subref*{Fig6c} show that chaotic LFF regime is recovered with Gaussian statistics as if the system had a single feedback with $\langle T \rangle$ down to the single feedback chaos case. 
In general similar chaos statistics occur for all Farey numbers $p/q$ where $p$ round trips in one cavity match $q$ roundtrips of the other one. The novelty to be emphasised here is that such quasi-periodicity in feedback times enters the laser dynamics to guarantee Gaussian chaos. 
When $q$ is big, the two cavities periodicity will only match in such long time, with respect to all relaxations times in the gain medium, so that all field coherence are washed out in between the compound cavities roundtrips. This restates the chaotic behavior again as in a single cavity. 

\subsection{Control of the Chaotic Statistics}

\begin{figure}[!h]
	\subfloat{
	\includegraphics[width=0.48\linewidth]{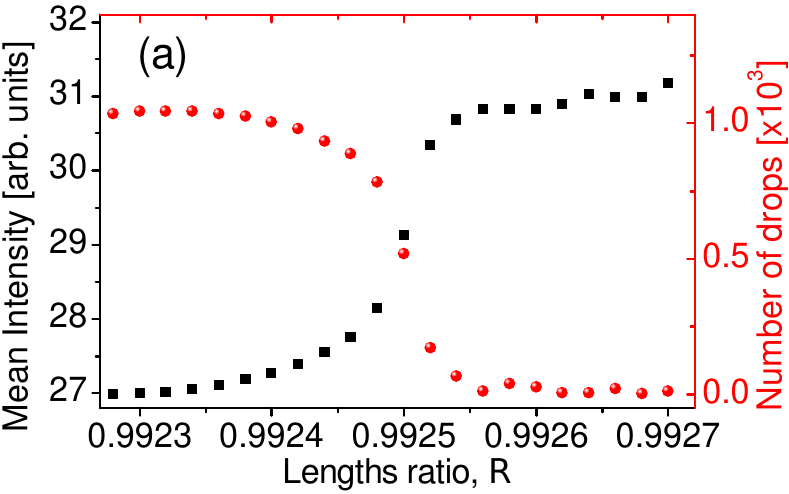}%[width=0.55\linewidth]
 	\label{Fig7a}}
	\subfloat{
   	\includegraphics[width=0.48\linewidth]{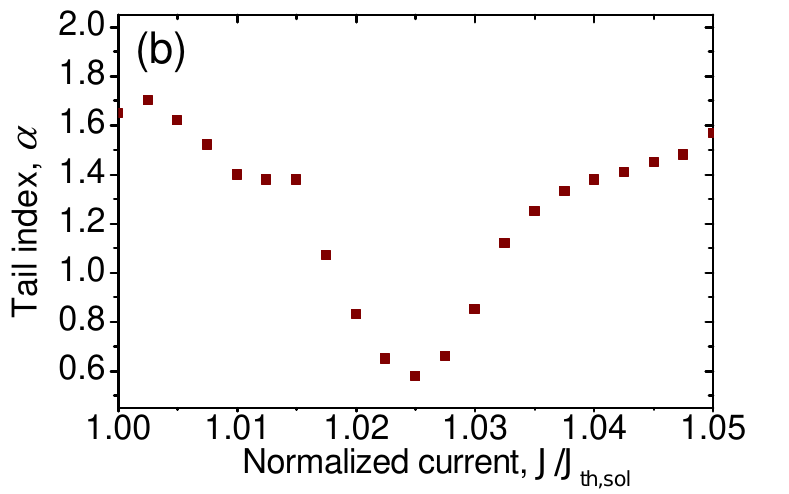}%[width=0.55\linewidth]
 	\label{Fig7b}}\\
	\subfloat{
   	\includegraphics[width=0.975\linewidth]{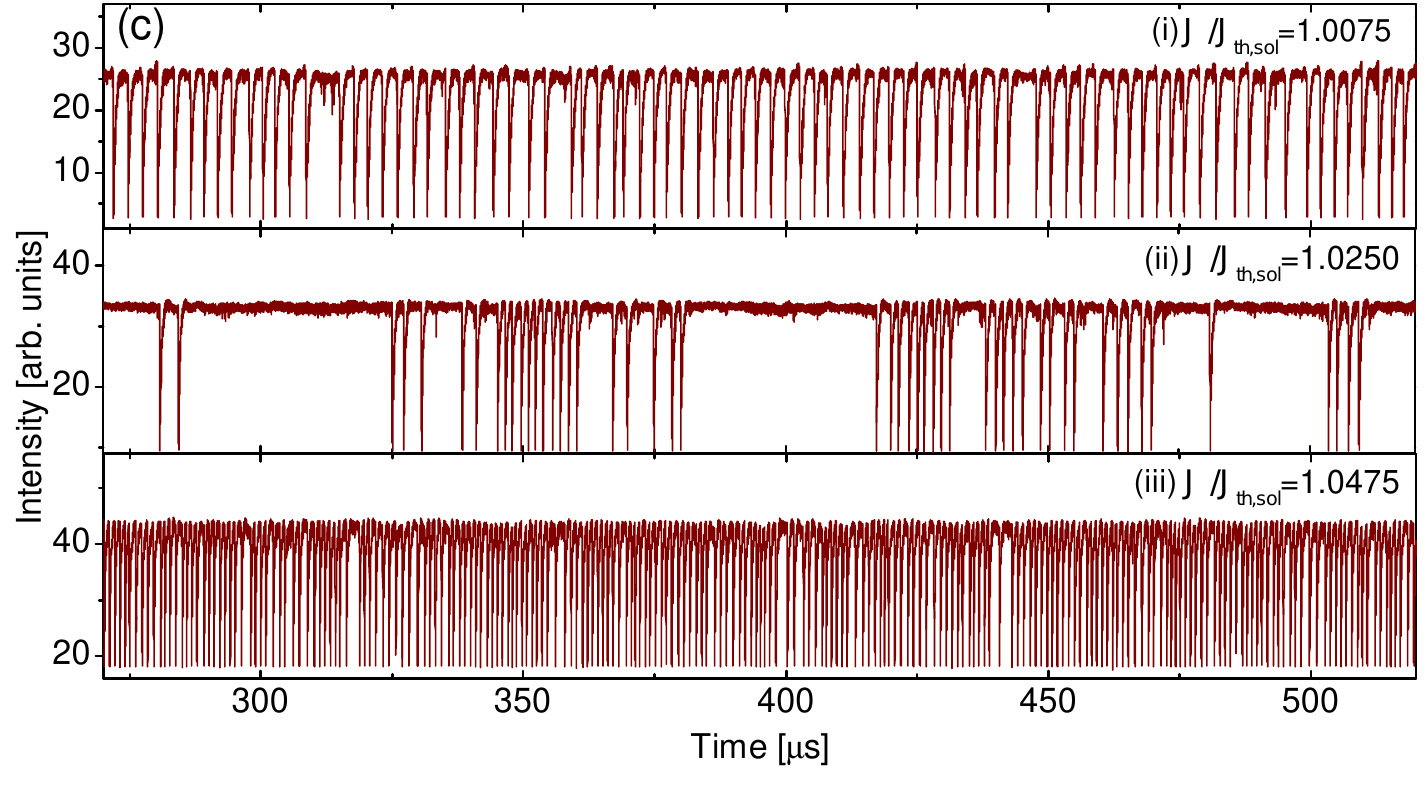}%[width=0.55\linewidth]
 	\label{Fig7c}}
\caption{Fine tuning of the chaos statistics. (a) Numerical calculation of the averaged output intensity (black squares online) and the number of LFF spikes (red balls online) for a transitions between different dynamical statistics near the ratio $R=1/1$. In scanning $R$ the step was $\delta R=0.00002$ which correspond to nearly 180 laser optical wavelength. (b) The long-tail statistical index $\alpha$ dependence on laser pump current  $J$ for the fixed ratio $R=0.9925$. The deep in the value of $\alpha$ from 1.7 to 0.6 and back to 1.7 reflects the change between Gaussian spiking chaos and long-tail with spike-bursting. (c) Intensity time series for points in (b): (i) $J/J_{th,sol}=1.0075$, (ii) $J/J_{th,sol}=1.0250$ and (iii) $J/J_{th,sol}=1.0475$.}
\label{Fig7}
\end{figure}

The long tail spike-bursting behavior is reached within these regions of transitions of ups and downs of $\langle T \rangle$. Considering that the change in $R$ within one of those $\langle T \rangle$ strong variations occurs for over thousands optical wavelengths, it suggests that without interferometric precision, one can adjust the system to have the desired statistics of spikes. Selecting the appropriate values for $R$ creates the possibility to have a system with probability distribution spikes with a tail index $\alpha$ chosen on demand. 
Figure \subref*{Fig7a} shows an even higher resolution numerical result for averaged output intensity of the laser and the number of LFF spikes of one of the chaos statistics transitions near the ratio $R=1/1$. 
%%%
%%%%% FIGURE 7 - SCAN J - CONTROL OF STATISTICS %%%%%
%%
It is important to notice that the averaged output intensity varies almost linearly with the mean inter-spike time $\langle T \rangle$. So, in Fig.~\subref*{Fig7a}, the growth of (black squares online) is matched with the decrease of the number of spikes drops (red balls online). The step used for $R$ scan in the calculations was $\delta R=0.00002$ which experimentally would correspond to a variation of 150 $\mu$ m in $L_2$. 
The optical wavelength in our lasers were 0.8 and 0.85 $\mu$m, so optical interferometry is secondary with respect to the observed dynamics as one tunes $R$. The laser suffers a transition from LFF chaos with Gaussian statistics to a regime with no intensity drops near $R=0.9925$. In this region the system presents the long-tail spike-bursting dynamics.

Other parameters are available to control the dynamics in addition to $R$ in the system. The easiest to manipulate is the pump current $J$. Envisaging control of the bursting statistics we calculated a fine tuning of the pump current maintaining $R$ fixed at the center of the transition presented in Fig.~\subref*{Fig7a}. This is shown if Fig.~\subref*{Fig7b}. The numerical range of $J$ variation was 5\% of the solitary threshold current $J_{th,sol}$ with steps of 0.25$\%$ of this value. A corresponding experimental situation of a laser with solitary threshold current of 20 mA would fall in the fairly controllable 50 $\mu$A precision in current tuning. 
The long-tail index $\alpha$ drops from 1.7 (near Gaussian distribution) to 0.6, which signals very long tail statistics, and then recovers back to near the Gaussian statistics value. Graphics (i), (ii) and (iii) in Fig.~\subref*{Fig7c} illustrate the nature of the described statistics with segments of intensity time series for different values of the pump current. This is a clear indication of the potential tuning of the current to control the chaotic bursting statistics.

\section{Conclusion and Outlook}

In this work, we have demonstrated experimentally, and confirmed with numerical solution of equation model, that a semiconductor laser with double feedback shows chaos statistics that can be controlled by realistic parameters variation. The chaotic behavior was manifested in the time intervals between power drops characteristic of the dynamics in such system.
The feedback times ratio and the pump current were shown to play a fundamental role in the transitions from chaos with Gaussian statistics to long-tail distributions characterized by spiking bursts. 
Our chaotic device with this possibility of control may find application in optical encrypted communications and in simulations of neural network systems. 

%\section*{Acknowledgments}

\begin{acknowledgments}
Work partially supported by Brazilian
Agencies Conselho Nacional de Pesquisa e Desenvolvimento (CNPq),
Funda\c{c}\~ao de Ci\^encia de Pernambuco (FACEPE) and
CAPES Programa PROEX 534/2018
and by a Brazil-France, CAPES-COFECUB project 456/04.
\end{acknowledgments}

\bibliography{References}

\begin{thebibliography}{42}%
\makeatletter
\providecommand \@ifxundefined [1]{%
 \@ifx{#1\undefined}
}%
\providecommand \@ifnum [1]{%
 \ifnum #1\expandafter \@firstoftwo
 \else \expandafter \@secondoftwo
 \fi
}%
\providecommand \@ifx [1]{%
 \ifx #1\expandafter \@firstoftwo
 \else \expandafter \@secondoftwo
 \fi
}%
\providecommand \natexlab [1]{#1}%
\providecommand \enquote  [1]{``#1''}%
\providecommand \bibnamefont  [1]{#1}%
\providecommand \bibfnamefont [1]{#1}%
\providecommand \citenamefont [1]{#1}%
\providecommand \href@noop [0]{\@secondoftwo}%
\providecommand \href [0]{\begingroup \@sanitize@url \@href}%
\providecommand \@href[1]{\@@startlink{#1}\@@href}%
\providecommand \@@href[1]{\endgroup#1\@@endlink}%
\providecommand \@sanitize@url [0]{\catcode `\\12\catcode `\$12\catcode
  `\&12\catcode `\#12\catcode `\^12\catcode `\_12\catcode `\%12\relax}%
\providecommand \@@startlink[1]{}%
\providecommand \@@endlink[0]{}%
\providecommand \url  [0]{\begingroup\@sanitize@url \@url }%
\providecommand \@url [1]{\endgroup\@href {#1}{\urlprefix }}%
\providecommand \urlprefix  [0]{URL }%
\providecommand \Eprint [0]{\href }%
\providecommand \doibase [0]{http://dx.doi.org/}%
\providecommand \selectlanguage [0]{\@gobble}%
\providecommand \bibinfo  [0]{\@secondoftwo}%
\providecommand \bibfield  [0]{\@secondoftwo}%
\providecommand \translation [1]{[#1]}%
\providecommand \BibitemOpen [0]{}%
\providecommand \bibitemStop [0]{}%
\providecommand \bibitemNoStop [0]{.\EOS\space}%
\providecommand \EOS [0]{\spacefactor3000\relax}%
\providecommand \BibitemShut  [1]{\csname bibitem#1\endcsname}%
\let\auto@bib@innerbib\@empty
%</preamble>
\bibitem [{\citenamefont {Ott}(2002)}]{Ott}%
  \BibitemOpen
  \bibfield  {author} {\bibinfo {author} {\bibfnamefont {E.}~\bibnamefont
  {Ott}},\ }\href {\doibase 10.1017/CBO9780511803260} {\emph {\bibinfo {title}
  {Chaos in Dynamical Systems}}},\ \bibinfo {edition} {2nd}\ ed.\ (\bibinfo
  {publisher} {Cambridge University Press},\ \bibinfo {year}
  {2002})\BibitemShut {NoStop}%
\bibitem [{\citenamefont {Sciamanna}\ and\ \citenamefont
  {Shore}(2015)}]{sciamanna-nature2015}%
  \BibitemOpen
  \bibfield  {author} {\bibinfo {author} {\bibfnamefont {M.}~\bibnamefont
  {Sciamanna}}\ and\ \bibinfo {author} {\bibfnamefont {K.~A.}\ \bibnamefont
  {Shore}},\ }\href {https://doi.org/10.1038/nphoton.2014.326} {\bibfield
  {journal} {\bibinfo  {journal} {Nature Photonics}\ }\textbf {\bibinfo
  {volume} {9}},\ \bibinfo {pages} {151 EP } (\bibinfo {year}
  {2015})}\BibitemShut {NoStop}%
\bibitem [{\citenamefont {Mackey}\ and\ \citenamefont
  {Glass}(1977)}]{mackey-glass1977-science}%
  \BibitemOpen
  \bibfield  {author} {\bibinfo {author} {\bibfnamefont {M.}~\bibnamefont
  {Mackey}}\ and\ \bibinfo {author} {\bibfnamefont {L.}~\bibnamefont {Glass}},\
  }\href {\doibase 10.1126/science.267326} {\bibfield  {journal} {\bibinfo
  {journal} {Science}\ }\textbf {\bibinfo {volume} {197}},\ \bibinfo {pages}
  {287} (\bibinfo {year} {1977})}\BibitemShut {NoStop}%
\bibitem [{\citenamefont {Rinzel}\ and\ \citenamefont
  {Ermentrout}(1989)}]{Rinzel1998}%
  \BibitemOpen
  \bibfield  {author} {\bibinfo {author} {\bibfnamefont {J.}~\bibnamefont
  {Rinzel}}\ and\ \bibinfo {author} {\bibfnamefont {G.}~\bibnamefont
  {Ermentrout}},\ }\enquote {\bibinfo {title} {Analysis of neural excitability
  and oscillations},}\ in\ \href@noop {} {\emph {\bibinfo {booktitle} {Methods
  in neuronal modeling}}},\ \bibinfo {editor} {edited by\ \bibinfo {editor}
  {\bibfnamefont {C.}~\bibnamefont {Koch}}\ and\ \bibinfo {editor}
  {\bibfnamefont {I.}~\bibnamefont {Segev}}}\ (\bibinfo  {publisher} {MIT
  Press},\ \bibinfo {year} {1989})\ pp.\ \bibinfo {pages}
  {135--169}\BibitemShut {NoStop}%
\bibitem [{\citenamefont {Izhikevich}(2000)}]{izki}%
  \BibitemOpen
  \bibfield  {author} {\bibinfo {author} {\bibfnamefont {E.}~\bibnamefont
  {Izhikevich}},\ }\href@noop {} {\bibfield  {journal} {\bibinfo  {journal}
  {International Journal of Bifurcation and Chaos in Applied Sciences and
  Engineering}\ }\textbf {\bibinfo {volume} {10}},\ \bibinfo {pages} {1171}
  (\bibinfo {year} {2000})}\BibitemShut {NoStop}%
\bibitem [{\citenamefont {Laing}\ \emph {et~al.}(2003)\citenamefont {Laing},
  \citenamefont {Doiron}, \citenamefont {Longtin}, \citenamefont {Noonan},
  \citenamefont {Turner},\ and\ \citenamefont {Maler}}]{Laing2003}%
  \BibitemOpen
  \bibfield  {author} {\bibinfo {author} {\bibfnamefont {C.~R.}\ \bibnamefont
  {Laing}}, \bibinfo {author} {\bibfnamefont {B.}~\bibnamefont {Doiron}},
  \bibinfo {author} {\bibfnamefont {A.}~\bibnamefont {Longtin}}, \bibinfo
  {author} {\bibfnamefont {L.}~\bibnamefont {Noonan}}, \bibinfo {author}
  {\bibfnamefont {R.~W.}\ \bibnamefont {Turner}}, \ and\ \bibinfo {author}
  {\bibfnamefont {L.}~\bibnamefont {Maler}},\ }\href {\doibase
  10.1023/A:1023269128622} {\bibfield  {journal} {\bibinfo  {journal} {Journal
  of Computational Neuroscience}\ }\textbf {\bibinfo {volume} {14}},\ \bibinfo
  {pages} {329} (\bibinfo {year} {2003})}\BibitemShut {NoStop}%
\bibitem [{\citenamefont {Krahe}\ and\ \citenamefont
  {Gabbiani}(2004)}]{NeuronBurstReview}%
  \BibitemOpen
  \bibfield  {author} {\bibinfo {author} {\bibfnamefont {R.}~\bibnamefont
  {Krahe}}\ and\ \bibinfo {author} {\bibfnamefont {F.}~\bibnamefont
  {Gabbiani}},\ }\href {https://doi.org/10.1038/nrn1296} {\bibfield  {journal}
  {\bibinfo  {journal} {Nature Reviews Neuroscience}\ }\textbf {\bibinfo
  {volume} {5}},\ \bibinfo {pages} {13 EP } (\bibinfo {year}
  {2004})}\BibitemShut {NoStop}%
\bibitem [{\citenamefont {Giudici}\ \emph {et~al.}(1997)\citenamefont
  {Giudici}, \citenamefont {Green}, \citenamefont {Giacomelli}, \citenamefont
  {Nespolo},\ and\ \citenamefont {R.~Tredicce}}]{giudici}%
  \BibitemOpen
  \bibfield  {author} {\bibinfo {author} {\bibfnamefont {M.}~\bibnamefont
  {Giudici}}, \bibinfo {author} {\bibfnamefont {C.}~\bibnamefont {Green}},
  \bibinfo {author} {\bibfnamefont {G.}~\bibnamefont {Giacomelli}}, \bibinfo
  {author} {\bibfnamefont {U.}~\bibnamefont {Nespolo}}, \ and\ \bibinfo
  {author} {\bibfnamefont {J.}~\bibnamefont {R.~Tredicce}},\ }\href {\doibase
  10.1103/PhysRevE.55.6414} {\bibfield  {journal} {\bibinfo  {journal}
  {Physical Review E - PHYS REV E}\ }\textbf {\bibinfo {volume} {55}},\
  \bibinfo {pages} {6414} (\bibinfo {year} {1997})}\BibitemShut {NoStop}%
\bibitem [{\citenamefont {Mos}\ \emph {et~al.}(2000)\citenamefont {Mos},
  \citenamefont {Hoppenbrouwers}, \citenamefont {Hill}, \citenamefont {Blum},
  \citenamefont {Schleipen},\ and\ \citenamefont {de~waardt}}]{mos}%
  \BibitemOpen
  \bibfield  {author} {\bibinfo {author} {\bibfnamefont {E.~C.}\ \bibnamefont
  {Mos}}, \bibinfo {author} {\bibfnamefont {J.~J.~L.}\ \bibnamefont
  {Hoppenbrouwers}}, \bibinfo {author} {\bibfnamefont {M.~T.}\ \bibnamefont
  {Hill}}, \bibinfo {author} {\bibfnamefont {M.~W.}\ \bibnamefont {Blum}},
  \bibinfo {author} {\bibfnamefont {J.~J. H.~B.}\ \bibnamefont {Schleipen}}, \
  and\ \bibinfo {author} {\bibfnamefont {H.}~\bibnamefont {de~waardt}},\ }\href
  {\doibase 10.1109/72.857778} {\bibfield  {journal} {\bibinfo  {journal} {IEEE
  Transactions on Neural Networks}\ }\textbf {\bibinfo {volume} {11}},\
  \bibinfo {pages} {988} (\bibinfo {year} {2000})}\BibitemShut {NoStop}%
\bibitem [{\citenamefont {Romariz}\ and\ \citenamefont
  {Wagner}(2007)}]{romariz}%
  \BibitemOpen
  \bibfield  {author} {\bibinfo {author} {\bibfnamefont {A.}~\bibnamefont
  {Romariz}}\ and\ \bibinfo {author} {\bibfnamefont {K.}~\bibnamefont
  {Wagner}},\ }\href {\doibase 10.1364/AO.46.004736} {\bibfield  {journal}
  {\bibinfo  {journal} {Applied optics}\ }\textbf {\bibinfo {volume} {46}},\
  \bibinfo {pages} {4736} (\bibinfo {year} {2007})}\BibitemShut {NoStop}%
\bibitem [{\citenamefont {Hurtado}\ \emph {et~al.}(2010)\citenamefont
  {Hurtado}, \citenamefont {Henning},\ and\ \citenamefont {Adams}}]{hurtado}%
  \BibitemOpen
  \bibfield  {author} {\bibinfo {author} {\bibfnamefont {A.}~\bibnamefont
  {Hurtado}}, \bibinfo {author} {\bibfnamefont {I.~D.}\ \bibnamefont
  {Henning}}, \ and\ \bibinfo {author} {\bibfnamefont {M.~J.}\ \bibnamefont
  {Adams}},\ }\href {\doibase 10.1364/OE.18.025170} {\bibfield  {journal}
  {\bibinfo  {journal} {Opt. Express}\ }\textbf {\bibinfo {volume} {18}},\
  \bibinfo {pages} {25170} (\bibinfo {year} {2010})}\BibitemShut {NoStop}%
\bibitem [{\citenamefont {Coomans}\ \emph {et~al.}(2011)\citenamefont
  {Coomans}, \citenamefont {Gelens}, \citenamefont {Beri}, \citenamefont
  {Danckaert},\ and\ \citenamefont {Van~der Sande}}]{coomans}%
  \BibitemOpen
  \bibfield  {author} {\bibinfo {author} {\bibfnamefont {W.}~\bibnamefont
  {Coomans}}, \bibinfo {author} {\bibfnamefont {L.}~\bibnamefont {Gelens}},
  \bibinfo {author} {\bibfnamefont {S.}~\bibnamefont {Beri}}, \bibinfo {author}
  {\bibfnamefont {J.}~\bibnamefont {Danckaert}}, \ and\ \bibinfo {author}
  {\bibfnamefont {G.}~\bibnamefont {Van~der Sande}},\ }\href {\doibase
  10.1103/PhysRevE.84.036209} {\bibfield  {journal} {\bibinfo  {journal}
  {Physical review. E, Statistical, nonlinear, and soft matter physics}\
  }\textbf {\bibinfo {volume} {84}},\ \bibinfo {pages} {036209} (\bibinfo
  {year} {2011})}\BibitemShut {NoStop}%
\bibitem [{\citenamefont {Selmi}\ \emph {et~al.}(2014)\citenamefont {Selmi},
  \citenamefont {Braive}, \citenamefont {Beaudoin}, \citenamefont {Sagnes},
  \citenamefont {Kuszelewicz},\ and\ \citenamefont {Barbay}}]{selmi}%
  \BibitemOpen
  \bibfield  {author} {\bibinfo {author} {\bibfnamefont {F.}~\bibnamefont
  {Selmi}}, \bibinfo {author} {\bibfnamefont {R.}~\bibnamefont {Braive}},
  \bibinfo {author} {\bibfnamefont {G.}~\bibnamefont {Beaudoin}}, \bibinfo
  {author} {\bibfnamefont {I.}~\bibnamefont {Sagnes}}, \bibinfo {author}
  {\bibfnamefont {R.}~\bibnamefont {Kuszelewicz}}, \ and\ \bibinfo {author}
  {\bibfnamefont {S.}~\bibnamefont {Barbay}},\ }\href {\doibase
  10.1103/PhysRevLett.112.183902} {\bibfield  {journal} {\bibinfo  {journal}
  {Phys. Rev. Lett.}\ }\textbf {\bibinfo {volume} {112}},\ \bibinfo {pages}
  {183902} (\bibinfo {year} {2014})}\BibitemShut {NoStop}%
\bibitem [{\citenamefont {M\'endez}\ \emph {et~al.}(2005)\citenamefont
  {M\'endez}, \citenamefont {Aliaga},\ and\ \citenamefont {Mindlin}}]{eguia2}%
  \BibitemOpen
  \bibfield  {author} {\bibinfo {author} {\bibfnamefont {J.~M.}\ \bibnamefont
  {M\'endez}}, \bibinfo {author} {\bibfnamefont {J.}~\bibnamefont {Aliaga}}, \
  and\ \bibinfo {author} {\bibfnamefont {G.~B.}\ \bibnamefont {Mindlin}},\
  }\href {\doibase 10.1103/PhysRevE.71.026231} {\bibfield  {journal} {\bibinfo
  {journal} {Phys. Rev. E}\ }\textbf {\bibinfo {volume} {71}},\ \bibinfo
  {pages} {026231} (\bibinfo {year} {2005})}\BibitemShut {NoStop}%
\bibitem [{\citenamefont {Martinez~Avila}\ \emph {et~al.}(2008)\citenamefont
  {Martinez~Avila}, \citenamefont {de~S.~Cavalcante},\ and\ \citenamefont
  {Rios~Leite}}]{jhonPRL2008}%
  \BibitemOpen
  \bibfield  {author} {\bibinfo {author} {\bibfnamefont {J.~F.}\ \bibnamefont
  {Martinez~Avila}}, \bibinfo {author} {\bibfnamefont {H.~L.~D.}\ \bibnamefont
  {de~S.~Cavalcante}}, \ and\ \bibinfo {author} {\bibfnamefont {J.~R.}\
  \bibnamefont {Rios~Leite}},\ }\href {\doibase 10.1103/PhysRevLett.100.044101}
  {\bibfield  {journal} {\bibinfo  {journal} {Phys. Rev. Lett.}\ }\textbf
  {\bibinfo {volume} {100}},\ \bibinfo {pages} {044101} (\bibinfo {year}
  {2008})}\BibitemShut {NoStop}%
\bibitem [{\citenamefont {Eguia}\ and\ \citenamefont {Mindlin}(1999)}]{eguia1}%
  \BibitemOpen
  \bibfield  {author} {\bibinfo {author} {\bibfnamefont {M.~C.}\ \bibnamefont
  {Eguia}}\ and\ \bibinfo {author} {\bibfnamefont {G.~B.}\ \bibnamefont
  {Mindlin}},\ }\href {\doibase 10.1103/PhysRevE.60.1551} {\bibfield  {journal}
  {\bibinfo  {journal} {Phys. Rev. E}\ }\textbf {\bibinfo {volume} {60}},\
  \bibinfo {pages} {1551} (\bibinfo {year} {1999})}\BibitemShut {NoStop}%
\bibitem [{\citenamefont {Kim}\ \emph {et~al.}(2015)\citenamefont {Kim},
  \citenamefont {Locquet}, \citenamefont {Choi},\ and\ \citenamefont
  {Citrin}}]{kim}%
  \BibitemOpen
  \bibfield  {author} {\bibinfo {author} {\bibfnamefont {B.}~\bibnamefont
  {Kim}}, \bibinfo {author} {\bibfnamefont {A.}~\bibnamefont {Locquet}},
  \bibinfo {author} {\bibfnamefont {D.}~\bibnamefont {Choi}}, \ and\ \bibinfo
  {author} {\bibfnamefont {D.~S.}\ \bibnamefont {Citrin}},\ }\href {\doibase
  10.1103/PhysRevA.91.061802} {\bibfield  {journal} {\bibinfo  {journal} {Phys.
  Rev. A}\ }\textbf {\bibinfo {volume} {91}},\ \bibinfo {pages} {061802}
  (\bibinfo {year} {2015})}\BibitemShut {NoStop}%
\bibitem [{\citenamefont {Masoller}\ and\ \citenamefont
  {Abraham}(1998)}]{massoler1}%
  \BibitemOpen
  \bibfield  {author} {\bibinfo {author} {\bibfnamefont {C.}~\bibnamefont
  {Masoller}}\ and\ \bibinfo {author} {\bibfnamefont {N.~B.}\ \bibnamefont
  {Abraham}},\ }\href {\doibase 10.1088/1355-5111/10/3/010} {\bibfield
  {journal} {\bibinfo  {journal} {Quantum and Semiclassical Optics: Journal of
  the European Optical Society Part B}\ }\textbf {\bibinfo {volume} {10}},\
  \bibinfo {pages} {519} (\bibinfo {year} {1998})}\BibitemShut {NoStop}%
\bibitem [{\citenamefont {Risch}\ and\ \citenamefont {Voumard}(1977)}]{Risch}%
  \BibitemOpen
  \bibfield  {author} {\bibinfo {author} {\bibfnamefont {C.}~\bibnamefont
  {Risch}}\ and\ \bibinfo {author} {\bibfnamefont {C.}~\bibnamefont
  {Voumard}},\ }\href {\doibase 10.1063/1.323922} {\bibfield  {journal}
  {\bibinfo  {journal} {Journal of Applied Physics}\ }\textbf {\bibinfo
  {volume} {48}},\ \bibinfo {pages} {2083} (\bibinfo {year} {1977})},\ \Eprint
  {http://arxiv.org/abs/https://doi.org/10.1063/1.323922}
  {https://doi.org/10.1063/1.323922} \BibitemShut {NoStop}%
\bibitem [{\citenamefont {Hohl}\ \emph {et~al.}(1995)\citenamefont {Hohl},
  \citenamefont {van~der Linden},\ and\ \citenamefont {Roy}}]{hohl}%
  \BibitemOpen
  \bibfield  {author} {\bibinfo {author} {\bibfnamefont {A.}~\bibnamefont
  {Hohl}}, \bibinfo {author} {\bibfnamefont {H.~J.~C.}\ \bibnamefont {van~der
  Linden}}, \ and\ \bibinfo {author} {\bibfnamefont {R.}~\bibnamefont {Roy}},\
  }\href {\doibase 10.1364/OL.20.002396} {\bibfield  {journal} {\bibinfo
  {journal} {Opt. Lett.}\ }\textbf {\bibinfo {volume} {20}},\ \bibinfo {pages}
  {2396} (\bibinfo {year} {1995})}\BibitemShut {NoStop}%
\bibitem [{\citenamefont {Sukow}\ \emph {et~al.}(1997)\citenamefont {Sukow},
  \citenamefont {Gardner},\ and\ \citenamefont {Gauthier}}]{sukow}%
  \BibitemOpen
  \bibfield  {author} {\bibinfo {author} {\bibfnamefont {D.~W.}\ \bibnamefont
  {Sukow}}, \bibinfo {author} {\bibfnamefont {J.~R.}\ \bibnamefont {Gardner}},
  \ and\ \bibinfo {author} {\bibfnamefont {D.~J.}\ \bibnamefont {Gauthier}},\
  }\href {\doibase 10.1103/PhysRevA.56.R3370} {\bibfield  {journal} {\bibinfo
  {journal} {Phys. Rev. A}\ }\textbf {\bibinfo {volume} {56}},\ \bibinfo
  {pages} {R3370} (\bibinfo {year} {1997})}\BibitemShut {NoStop}%
\bibitem [{\citenamefont {Mulet}\ and\ \citenamefont {Mirasso}(1999)}]{mulet}%
  \BibitemOpen
  \bibfield  {author} {\bibinfo {author} {\bibfnamefont {J.}~\bibnamefont
  {Mulet}}\ and\ \bibinfo {author} {\bibfnamefont {C.~R.}\ \bibnamefont
  {Mirasso}},\ }\href {\doibase 10.1103/PhysRevE.59.5400} {\bibfield  {journal}
  {\bibinfo  {journal} {Phys. Rev. E}\ }\textbf {\bibinfo {volume} {59}},\
  \bibinfo {pages} {5400} (\bibinfo {year} {1999})}\BibitemShut {NoStop}%
\bibitem [{\citenamefont {Aragoneses}\ \emph {et~al.}(2013)\citenamefont
  {Aragoneses}, \citenamefont {Rubido}, \citenamefont {Tiana-Alsina},
  \citenamefont {Torrent},\ and\ \citenamefont {Masoller}}]{massoler2014}%
  \BibitemOpen
  \bibfield  {author} {\bibinfo {author} {\bibfnamefont {A.}~\bibnamefont
  {Aragoneses}}, \bibinfo {author} {\bibfnamefont {N.}~\bibnamefont {Rubido}},
  \bibinfo {author} {\bibfnamefont {J.}~\bibnamefont {Tiana-Alsina}}, \bibinfo
  {author} {\bibfnamefont {M.~C.}\ \bibnamefont {Torrent}}, \ and\ \bibinfo
  {author} {\bibfnamefont {C.}~\bibnamefont {Masoller}},\ }\href
  {https://doi.org/10.1038/srep01778} {\bibfield  {journal} {\bibinfo
  {journal} {Scientific Reports}\ }\textbf {\bibinfo {volume} {3}},\ \bibinfo
  {pages} {1778 EP } (\bibinfo {year} {2013})}\BibitemShut {NoStop}%
\bibitem [{\citenamefont {Rogister}\ \emph {et~al.}(1999)\citenamefont
  {Rogister}, \citenamefont {M\'{e}gret}, \citenamefont {Deparis},
  \citenamefont {Blondel},\ and\ \citenamefont {Erneux}}]{rogisternum}%
  \BibitemOpen
  \bibfield  {author} {\bibinfo {author} {\bibfnamefont {F.}~\bibnamefont
  {Rogister}}, \bibinfo {author} {\bibfnamefont {P.}~\bibnamefont
  {M\'{e}gret}}, \bibinfo {author} {\bibfnamefont {O.}~\bibnamefont {Deparis}},
  \bibinfo {author} {\bibfnamefont {M.}~\bibnamefont {Blondel}}, \ and\
  \bibinfo {author} {\bibfnamefont {T.}~\bibnamefont {Erneux}},\ }\href
  {\doibase 10.1364/OL.24.001218} {\bibfield  {journal} {\bibinfo  {journal}
  {Opt. Lett.}\ }\textbf {\bibinfo {volume} {24}},\ \bibinfo {pages} {1218}
  (\bibinfo {year} {1999})}\BibitemShut {NoStop}%
\bibitem [{\citenamefont {Rogister}\ \emph {et~al.}(2000)\citenamefont
  {Rogister}, \citenamefont {Sukow}, \citenamefont {Gavrielides}, \citenamefont
  {M\'{e}gret}, \citenamefont {Deparis},\ and\ \citenamefont
  {Blondel}}]{rogisterexp}%
  \BibitemOpen
  \bibfield  {author} {\bibinfo {author} {\bibfnamefont {F.}~\bibnamefont
  {Rogister}}, \bibinfo {author} {\bibfnamefont {D.~W.}\ \bibnamefont {Sukow}},
  \bibinfo {author} {\bibfnamefont {A.}~\bibnamefont {Gavrielides}}, \bibinfo
  {author} {\bibfnamefont {P.}~\bibnamefont {M\'{e}gret}}, \bibinfo {author}
  {\bibfnamefont {O.}~\bibnamefont {Deparis}}, \ and\ \bibinfo {author}
  {\bibfnamefont {M.}~\bibnamefont {Blondel}},\ }\href {\doibase
  10.1364/OL.25.000808} {\bibfield  {journal} {\bibinfo  {journal} {Opt.
  Lett.}\ }\textbf {\bibinfo {volume} {25}},\ \bibinfo {pages} {808} (\bibinfo
  {year} {2000})}\BibitemShut {NoStop}%
\bibitem [{\citenamefont {Ruiz-Oliveras}\ and\ \citenamefont
  {Pisarchik}(2006)}]{pisarchik}%
  \BibitemOpen
  \bibfield  {author} {\bibinfo {author} {\bibfnamefont {F.~R.}\ \bibnamefont
  {Ruiz-Oliveras}}\ and\ \bibinfo {author} {\bibfnamefont {A.~N.}\ \bibnamefont
  {Pisarchik}},\ }\href {\doibase 10.1364/OE.14.012859} {\bibfield  {journal}
  {\bibinfo  {journal} {Opt. Express}\ }\textbf {\bibinfo {volume} {14}},\
  \bibinfo {pages} {12859} (\bibinfo {year} {2006})}\BibitemShut {NoStop}%
\bibitem [{\citenamefont {Wu}\ \emph {et~al.}(2009)\citenamefont {Wu},
  \citenamefont {Wu},\ and\ \citenamefont {Xia}}]{jia}%
  \BibitemOpen
  \bibfield  {author} {\bibinfo {author} {\bibfnamefont {J.-G.}\ \bibnamefont
  {Wu}}, \bibinfo {author} {\bibfnamefont {Z.-M.}\ \bibnamefont {Wu}}, \ and\
  \bibinfo {author} {\bibfnamefont {G.-Q.}\ \bibnamefont {Xia}},\ }\href
  {\doibase https://doi.org/10.1016/j.physleta.2009.10.039} {\bibfield
  {journal} {\bibinfo  {journal} {Physics Letters A}\ }\textbf {\bibinfo
  {volume} {374}},\ \bibinfo {pages} {173 } (\bibinfo {year}
  {2009})}\BibitemShut {NoStop}%
\bibitem [{\citenamefont {Fischer}\ \emph {et~al.}(1994)\citenamefont
  {Fischer}, \citenamefont {Hess}, \citenamefont {Els\"a\ensuremath{\beta}er},\
  and\ \citenamefont {G\"obel}}]{fisher}%
  \BibitemOpen
  \bibfield  {author} {\bibinfo {author} {\bibfnamefont {I.}~\bibnamefont
  {Fischer}}, \bibinfo {author} {\bibfnamefont {O.}~\bibnamefont {Hess}},
  \bibinfo {author} {\bibfnamefont {W.}~\bibnamefont
  {Els\"a\ensuremath{\beta}er}}, \ and\ \bibinfo {author} {\bibfnamefont
  {E.}~\bibnamefont {G\"obel}},\ }\href {\doibase 10.1103/PhysRevLett.73.2188}
  {\bibfield  {journal} {\bibinfo  {journal} {Phys. Rev. Lett.}\ }\textbf
  {\bibinfo {volume} {73}},\ \bibinfo {pages} {2188} (\bibinfo {year}
  {1994})}\BibitemShut {NoStop}%
\bibitem [{\citenamefont {Liu}\ and\ \citenamefont {Ohtsubo}(1997)}]{liu}%
  \BibitemOpen
  \bibfield  {author} {\bibinfo {author} {\bibfnamefont {Y.}~\bibnamefont
  {Liu}}\ and\ \bibinfo {author} {\bibfnamefont {J.}~\bibnamefont {Ohtsubo}},\
  }\href {\doibase 10.1109/3.594879} {\bibfield  {journal} {\bibinfo  {journal}
  {IEEE Journal of Quantum Electronics}\ }\textbf {\bibinfo {volume} {33}},\
  \bibinfo {pages} {1163} (\bibinfo {year} {1997})}\BibitemShut {NoStop}%
\bibitem [{\citenamefont {T\"obbens}\ and\ \citenamefont
  {Parlitz}(2008)}]{tobens}%
  \BibitemOpen
  \bibfield  {author} {\bibinfo {author} {\bibfnamefont {A.}~\bibnamefont
  {T\"obbens}}\ and\ \bibinfo {author} {\bibfnamefont {U.}~\bibnamefont
  {Parlitz}},\ }\href {\doibase 10.1103/PhysRevE.78.016210} {\bibfield
  {journal} {\bibinfo  {journal} {Phys. Rev. E}\ }\textbf {\bibinfo {volume}
  {78}},\ \bibinfo {pages} {016210} (\bibinfo {year} {2008})}\BibitemShut
  {NoStop}%
\bibitem [{\citenamefont {Lang}\ and\ \citenamefont {Kobayashi}(1980)}]{LK}%
  \BibitemOpen
  \bibfield  {author} {\bibinfo {author} {\bibfnamefont {R.}~\bibnamefont
  {Lang}}\ and\ \bibinfo {author} {\bibfnamefont {K.}~\bibnamefont
  {Kobayashi}},\ }\href {\doibase 10.1109/JQE.1980.1070479} {\bibfield
  {journal} {\bibinfo  {journal} {IEEE Journal of Quantum Electronics}\
  }\textbf {\bibinfo {volume} {16}},\ \bibinfo {pages} {347} (\bibinfo {year}
  {1980})}\BibitemShut {NoStop}%
\bibitem [{\citenamefont {Nolan}(2018)}]{LevyDistributions}%
  \BibitemOpen
  \bibfield  {author} {\bibinfo {author} {\bibfnamefont {J.~P.}\ \bibnamefont
  {Nolan}},\ }\href@noop {} {\emph {\bibinfo {title} {Stable Distributions -
  Models for Heavy Tailed Data}}}\ (\bibinfo  {publisher} {Birkhauser},\
  \bibinfo {address} {Boston},\ \bibinfo {year} {2018})\ \bibinfo {note} {in
  progress, Chapter 1 online at
  http://fs2.american.edu/jpnolan/www/stable/stable.html}\BibitemShut {NoStop}%
\bibitem [{\citenamefont {McCulloch}(1986)}]{McCulloch}%
  \BibitemOpen
  \bibfield  {author} {\bibinfo {author} {\bibfnamefont {J.~H.}\ \bibnamefont
  {McCulloch}},\ }\href {\doibase 10.1080/03610918608812563} {\bibfield
  {journal} {\bibinfo  {journal} {Communications in Statistics - Simulation and
  Computation}\ }\textbf {\bibinfo {volume} {15}},\ \bibinfo {pages} {1109}
  (\bibinfo {year} {1986})}\BibitemShut {NoStop}%
\bibitem [{\citenamefont {Nolan}()}]{NOLAN}%
  \BibitemOpen
  \bibfield  {author} {\bibinfo {author} {\bibfnamefont {J.~P.}\ \bibnamefont
  {Nolan}},\ }\href@noop {} {\enquote {\bibinfo {title} {Computer program for
  parameters of stable distributions},}\ }\BibitemShut {NoStop}%
\bibitem [{\citenamefont {Heil}\ \emph {et~al.}(2001)\citenamefont {Heil},
  \citenamefont {Fischer}, \citenamefont {Els\"a\ss{}er},\ and\ \citenamefont
  {Gavrielides}}]{heilShortCavity}%
  \BibitemOpen
  \bibfield  {author} {\bibinfo {author} {\bibfnamefont {T.}~\bibnamefont
  {Heil}}, \bibinfo {author} {\bibfnamefont {I.}~\bibnamefont {Fischer}},
  \bibinfo {author} {\bibfnamefont {W.}~\bibnamefont {Els\"a\ss{}er}}, \ and\
  \bibinfo {author} {\bibfnamefont {A.}~\bibnamefont {Gavrielides}},\ }\href
  {\doibase 10.1103/PhysRevLett.87.243901} {\bibfield  {journal} {\bibinfo
  {journal} {Phys. Rev. Lett.}\ }\textbf {\bibinfo {volume} {87}},\ \bibinfo
  {pages} {243901} (\bibinfo {year} {2001})}\BibitemShut {NoStop}%
\bibitem [{\citenamefont {Xu}\ \emph {et~al.}(1990)\citenamefont {Xu},
  \citenamefont {Dai}, \citenamefont {Yang}, \citenamefont {Zhang},\ and\
  \citenamefont {Zhang}}]{Xu-PRA}%
  \BibitemOpen
  \bibfield  {author} {\bibinfo {author} {\bibfnamefont {G.}~\bibnamefont
  {Xu}}, \bibinfo {author} {\bibfnamefont {J.-h.}\ \bibnamefont {Dai}},
  \bibinfo {author} {\bibfnamefont {S.-p.}\ \bibnamefont {Yang}}, \bibinfo
  {author} {\bibfnamefont {F.-l.}\ \bibnamefont {Zhang}}, \ and\ \bibinfo
  {author} {\bibfnamefont {H.-j.}\ \bibnamefont {Zhang}},\ }\href {\doibase
  10.1103/PhysRevA.42.4269} {\bibfield  {journal} {\bibinfo  {journal} {Phys.
  Rev. A}\ }\textbf {\bibinfo {volume} {42}},\ \bibinfo {pages} {4269}
  (\bibinfo {year} {1990})}\BibitemShut {NoStop}%
\bibitem [{\citenamefont {Sano}(1994)}]{sano}%
  \BibitemOpen
  \bibfield  {author} {\bibinfo {author} {\bibfnamefont {T.}~\bibnamefont
  {Sano}},\ }\href {\doibase 10.1103/PhysRevA.50.2719} {\bibfield  {journal}
  {\bibinfo  {journal} {Phys. Rev. A}\ }\textbf {\bibinfo {volume} {50}},\
  \bibinfo {pages} {2719} (\bibinfo {year} {1994})}\BibitemShut {NoStop}%
\bibitem [{\citenamefont {Zamora-Munt}\ \emph {et~al.}(2010)\citenamefont
  {Zamora-Munt}, \citenamefont {Masoller},\ and\ \citenamefont
  {Garc\'{\i}a-Ojalvo}}]{masoller3}%
  \BibitemOpen
  \bibfield  {author} {\bibinfo {author} {\bibfnamefont {J.}~\bibnamefont
  {Zamora-Munt}}, \bibinfo {author} {\bibfnamefont {C.}~\bibnamefont
  {Masoller}}, \ and\ \bibinfo {author} {\bibfnamefont {J.}~\bibnamefont
  {Garc\'{\i}a-Ojalvo}},\ }\href {\doibase 10.1103/PhysRevA.81.033820}
  {\bibfield  {journal} {\bibinfo  {journal} {Phys. Rev. A}\ }\textbf {\bibinfo
  {volume} {81}},\ \bibinfo {pages} {033820} (\bibinfo {year}
  {2010})}\BibitemShut {NoStop}%
\bibitem [{\citenamefont {Sacher}\ \emph {et~al.}(1989)\citenamefont {Sacher},
  \citenamefont {Els\"asser},\ and\ \citenamefont {G\"obel}}]{sacher}%
  \BibitemOpen
  \bibfield  {author} {\bibinfo {author} {\bibfnamefont {J.}~\bibnamefont
  {Sacher}}, \bibinfo {author} {\bibfnamefont {W.}~\bibnamefont {Els\"asser}},
  \ and\ \bibinfo {author} {\bibfnamefont {E.~O.}\ \bibnamefont {G\"obel}},\
  }\href {\doibase 10.1103/PhysRevLett.63.2224} {\bibfield  {journal} {\bibinfo
   {journal} {Phys. Rev. Lett.}\ }\textbf {\bibinfo {volume} {63}},\ \bibinfo
  {pages} {2224} (\bibinfo {year} {1989})}\BibitemShut {NoStop}%
\bibitem [{\citenamefont {Ikeda}\ and\ \citenamefont
  {Mizuno}(1984)}]{IkedaPRL1985}%
  \BibitemOpen
  \bibfield  {author} {\bibinfo {author} {\bibfnamefont {K.}~\bibnamefont
  {Ikeda}}\ and\ \bibinfo {author} {\bibfnamefont {M.}~\bibnamefont {Mizuno}},\
  }\href {\doibase 10.1103/PhysRevLett.53.1340} {\bibfield  {journal} {\bibinfo
   {journal} {Phys. Rev. Lett.}\ }\textbf {\bibinfo {volume} {53}},\ \bibinfo
  {pages} {1340} (\bibinfo {year} {1984})}\BibitemShut {NoStop}%
\bibitem [{\citenamefont {Ikeda}\ and\ \citenamefont
  {Mizuno}(1985)}]{ikedaIEEE1985}%
  \BibitemOpen
  \bibfield  {author} {\bibinfo {author} {\bibfnamefont {K.}~\bibnamefont
  {Ikeda}}\ and\ \bibinfo {author} {\bibfnamefont {M.}~\bibnamefont {Mizuno}},\
  }\href {\doibase 10.1109/JQE.1985.1072824} {\bibfield  {journal} {\bibinfo
  {journal} {IEEE Journal of Quantum Electronics}\ }\textbf {\bibinfo {volume}
  {21}},\ \bibinfo {pages} {1429} (\bibinfo {year} {1985})}\BibitemShut
  {NoStop}%
\bibitem [{\citenamefont {Mizuno}\ and\ \citenamefont {Ikeda}(1989)}]{mizuno}%
  \BibitemOpen
  \bibfield  {author} {\bibinfo {author} {\bibfnamefont {M.}~\bibnamefont
  {Mizuno}}\ and\ \bibinfo {author} {\bibfnamefont {K.}~\bibnamefont {Ikeda}},\
  }\href {\doibase https://doi.org/10.1016/0167-2789(89)90088-2} {\bibfield
  {journal} {\bibinfo  {journal} {Physica D: Nonlinear Phenomena}\ }\textbf
  {\bibinfo {volume} {36}},\ \bibinfo {pages} {327 } (\bibinfo {year}
  {1989})}\BibitemShut {NoStop}%
\end{thebibliography}%

\end{document}